\documentclass[aps,prb,preprint,showpacs,superscriptaddress,longbibliography]{revtex4-1}

\usepackage{amsmath,amsfonts}
\usepackage{graphicx}
\usepackage{multirow}

\usepackage{dcolumn}
\usepackage{longtable}
\usepackage{color}
\usepackage{setspace}

\begin{document}

\title{
    Spin-orbital Entangled Molecular $\boldsymbol{j}_{\rm eff}$ States \\
    in Lacunar Spinel Compounds
}

\author{Heung-Sik Kim}
\affiliation{Department of Physics, Korean Advanced Institute of Science and Technology,
Daejun 305-701, Korea}

\author{Jino Im}
\affiliation{Department of Physics and Astronomy, Northwestern University, Evanston,
Illinois 60208, USA}

\author{Myung Joon Han}
\affiliation{Department of Physics, Korean Advanced Institute of Science and Technology,
Daejun 305-701, Korea}
\affiliation{KAIST Institute for the NanoCentury, Korean Advanced Institute of Science and Technology,
Daejun 305-701, Korea}

\author{Hosub Jin}
\email{Correspondence: jinhs76@snu.ac.kr}
\affiliation{Center for Correlated Electron Systems, Institute for Basic Science (IBS), Seoul 151-747, Korea}
\affiliation{Department of Physics and Astronomy, Seoul National University, Seoul 151-747, Korea}

\maketitle

{\bf
The entanglement of the spin and orbital degrees of freedom through the spin-orbit coupling has been
actively studied in condensed matter physics. In several iridium-oxide systems, the spin-orbital
entangled state, identified by the effective angular momentum \emph{j}$_{\mathrm{eff}}$, can host
novel quantum phases with the help of electron correlations. Here we show that a series of lacunar
spinel compounds, Ga\emph{M}$_4$\emph{X}$_8$ (\emph{M} = Nb, Mo, Ta, and W and \emph{X} = S, Se, and Te),
gives rise to a molecular \emph{j}$_{\mathrm{eff}}$ state as a new spin-orbital composite
on which the low energy effective Hamiltonian is based. A wide range of electron correlations is
accessible by tuning the bandwidth under external and/or chemical pressure, enabling us to investigate
the interesting cooperation between spin-orbit coupling and electron correlations. As illustrative
examples, a two-dimensional topological insulating phase and an anisotropic spin Hamiltonian are
investigated in the weak and strong coupling regimes, respectively. Our finding can provide an ideal
platform for exploring \emph{j}$_{\mathrm{eff}}$ physics and the resulting emergent phenomena.
}

Spin-orbit coupling (SOC) is a manifestation of Einstein's theory of relativity in condensed matter
systems. Recently, SOC has attracted a great deal of attention since it is a main ingredient for spintronics
applications~\cite{Datta,Pesin:SHE}, induces novel quantum phases~\cite{KM,QAH:exp}, and generates new
particles and elementary excitations~\cite{Qi:monopole,Fu:SC}. Moreover, when incorporated with electron
correlations, SOC can give rise to even more fascinating phenomena~\cite{Pesin:TMI,Review:jeff}. In the
iridium oxide family, where the IrO$_6$ octahedron is the essential building block, various quantum phases
have been predicted or verified according to the electron correlation strength on top of the large SOC
of the Ir 5$d$ $t_{\rm 2g}$ orbital: topological band insulator for weak coupling~\cite{Shitade:NIO,Guo:TI},
Weyl semi-metal, axion insulator, non-Fermi liquid, and TI$^*$ phases for intermediate coupling~\cite{Wan:WS,Go:WS,Axion,Moon_LAB,TIstar},
and topological Mott insulator and quantum spin liquid phases for strong coupling~\cite{Okamoto,Pesin:TMI,Chaloupka:KH}.

Emergence of the spin-orbital entangled $j_{\rm eff}$ states induced by SOC~\cite{Kim:PRL,Kim:Science}
is the key feature to host all the above phases, yet the existence of such states is limited to a
small number of iridate compounds only. Here, the series of lacunar spinel compounds~\cite{Pocha:CM,Pocha:JACS},
Ga$M_4X_8$, where early 4$d$ or 5$d$ transition metal atoms occupy the $M$-site, are found to provide
the molecular form of the $j_{\rm eff}$ basis in their low energy electronic structures. The idealness
of the molecular $j_{\rm eff}$ state is guaranteed by the formation of the $M_4$ metal cluster and the
large SOC. Combined with the ability to control the electron correlation from the weak to strong coupling
limit, the lacunar spinels can manifest themselves as the best candidates to demonstrate this so-called
$j_{\rm eff}$ physics.

\textbf{\large{Results}}

\begin{figure}
  \centering
  \includegraphics[width=0.6\textwidth]{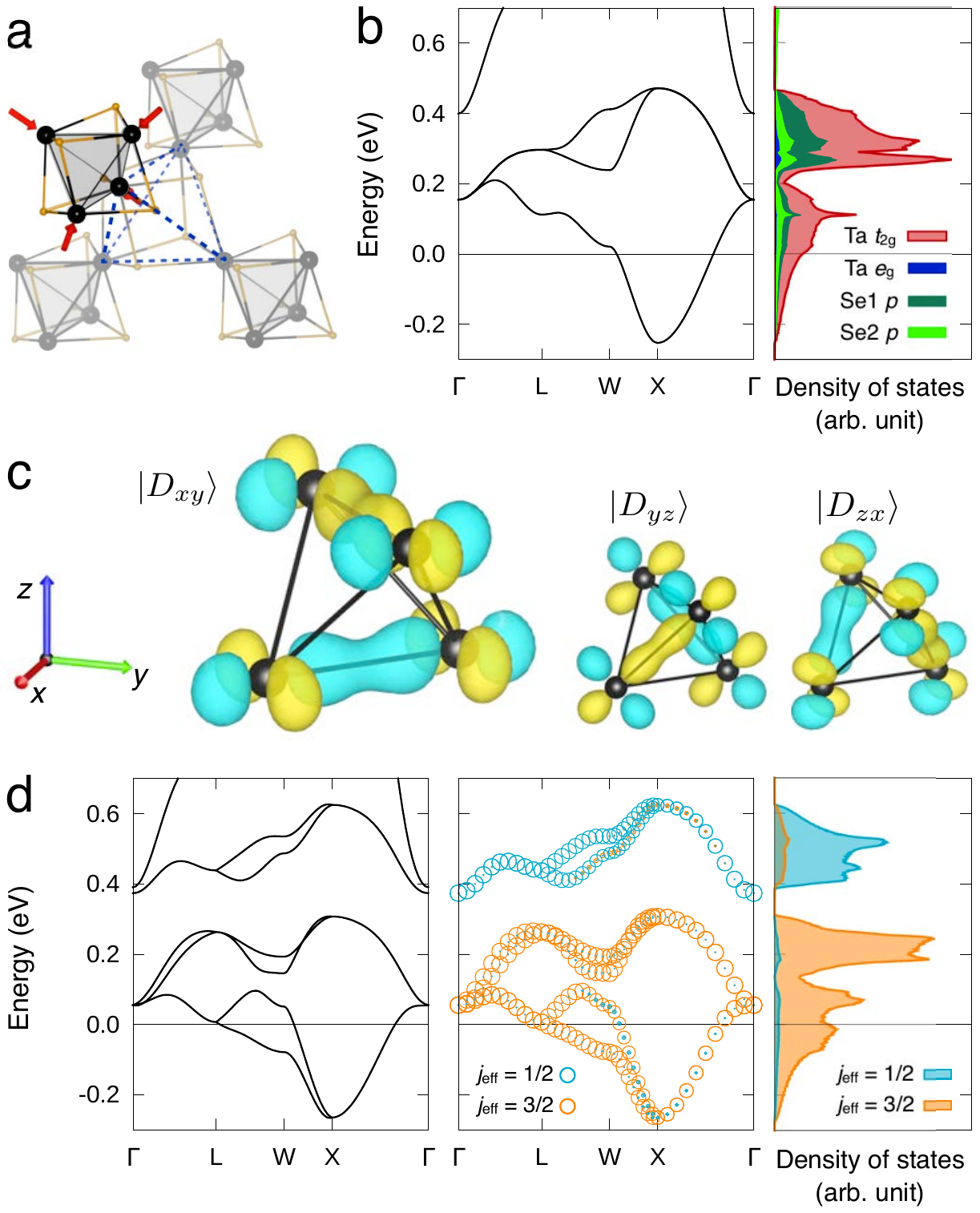}
  \caption{
        }
  \label{fig:atomdisp}
\end{figure}

\textbf{Formation of the molecular $\boldsymbol{j}_{\rm eff}$ states in GaTa$_4$Se$_8$.}
The chemical formula and crystal structure of the Ga$M_4X_8$ lacunar spinels are easily deduced from the spinel
with half-deficient Ga atoms, {\it i.e.} Ga$_{0.5}M_2X_4$. Due to the half-removal of the Ga atoms, the
transition metal atoms are strongly distorted into the tetrahedral center as denoted by the red arrows
in Fig.~\ref{fig:atomdisp} {\bf a}, and a tetramerized $M_4$ cluster appears. The $M_4$ cluster yields
a short intra-cluster $M$-$M$ distance, naturally inducing the molecular states residing on the cluster
as basic building blocks for the low energy electronic structure. On the other hand, the large inter-cluster
distance results in a weak inter-cluster bonding and a narrow bandwidth of the molecular states.

As a representative example of the lacunar spinels, we investigate the electronic structure of
GaTa$_4$Se$_8$ (Fig.~\ref{fig:atomdisp} {\bf b}-{\bf d}). Figure~\ref{fig:atomdisp} {\bf b}
shows the band structure and the projected density of states (PDOS) of GaTa$_4$Se$_8$ in the absence of SOC.
In consistency with previous studies~\cite{Pocha:JACS,Camjayi,TaPhuoc}, the triply degenerate molecular $t_2$
bands occupied by one electron are located near the Fermi level with a small bandwidth of $\sim$0.75~eV.
(See Supplementary Note 1, Supplementary Figure 1, and Supplementary Table 1
for details on the molecular $t_2$ effective Hamiltonian.)
As shown in the PDOS plot, the molecular $t_2$ bands are dominated by Ta $t_{\rm 2g}$ orbital components;
the small admixture of Se 5$p$ and the strong tetramerization imply that the molecular $t_2$ states
consist of direct bonding between Ta $t_{\rm 2g}$ states.

The molecular nature of the low-energy electronic structure can be visualized by adopting the maximally
localized Wannier function scheme~\cite{MLWF,MLWF:DISE}. The three molecular $t_2$ Wannier functions
depicted in Fig.~\ref{fig:atomdisp} {\bf c} read
\begin{equation}
\vert D_\alpha \rangle = \frac{1}{2} \sum^4_{i=1} \vert d^{~i}_\alpha \rangle ~~~ (\alpha = xy,yz,zx),
\label{eq:mt2}
\end{equation}
where $D_\alpha$ and $d_\alpha$ denote the molecular $t_2$ and atomic $t_{\rm 2g}$ states, respectively,
and $i$ is a site index indicating the four corners of the $M_4$ cluster. Each $D_{\alpha}$ originates
from a $\sigma$-type strong bonding between the constituent $t_{\rm 2g}$ orbitals in the $M_4$ cluster.
Owing to the exact correspondence between the molecular $t_2$ and the atomic $t_{\rm 2g}$ states, as
revealed in Eq.~\ref{eq:mt2}, the molecular $t_2$ triplet carries the same effective orbital angular
momentum $l_{\rm eff}$ = 1 as the atomic $t_{\rm 2g}$ orbital~\cite{Kim:PRL}. By virtue of SOC, the
$l_{\rm eff}$ = 1 states are entangled with the $s$ = 1/2 spin, and two multiplets designated by the
effective total angular momentum $j_{\rm eff}$ = 1/2 and 3/2 emerge. The band structure and PDOS of
GaTa$_4$Se$_8$ in the presence of SOC verify the above $j_{\rm eff}$ picture (Fig.~\ref{fig:atomdisp} {\bf d});
the molecular $t_2$ bands split into upper $j_{\rm eff}$ = 1/2 and lower $j_{\rm eff}$ = 3/2 bands.
The separation between the two $j_{\rm eff}$ subbands is almost perfect owing to the large SOC of the
Ta atoms as well as the small bandwidth of the molecular $t_2$ band. An alternative confirmation of
the $j_{\rm eff}$ picture can also be given by constructing the Wannier function from each of the $j_{\rm eff}$
subbands, which shows a 99\% agreement with the ideal molecular $j_{\rm eff}$ states. 
(See Supplementary Figure 2.)
Consequently, the electronic structure of GaTa$_4$Se$_8$ can be labeled as a quarter-filled $j_{\rm eff}$ = 3/2
system on a face-centered cubic lattice.

\begin{figure}
  \centering
  \includegraphics[width=1.05\textwidth]{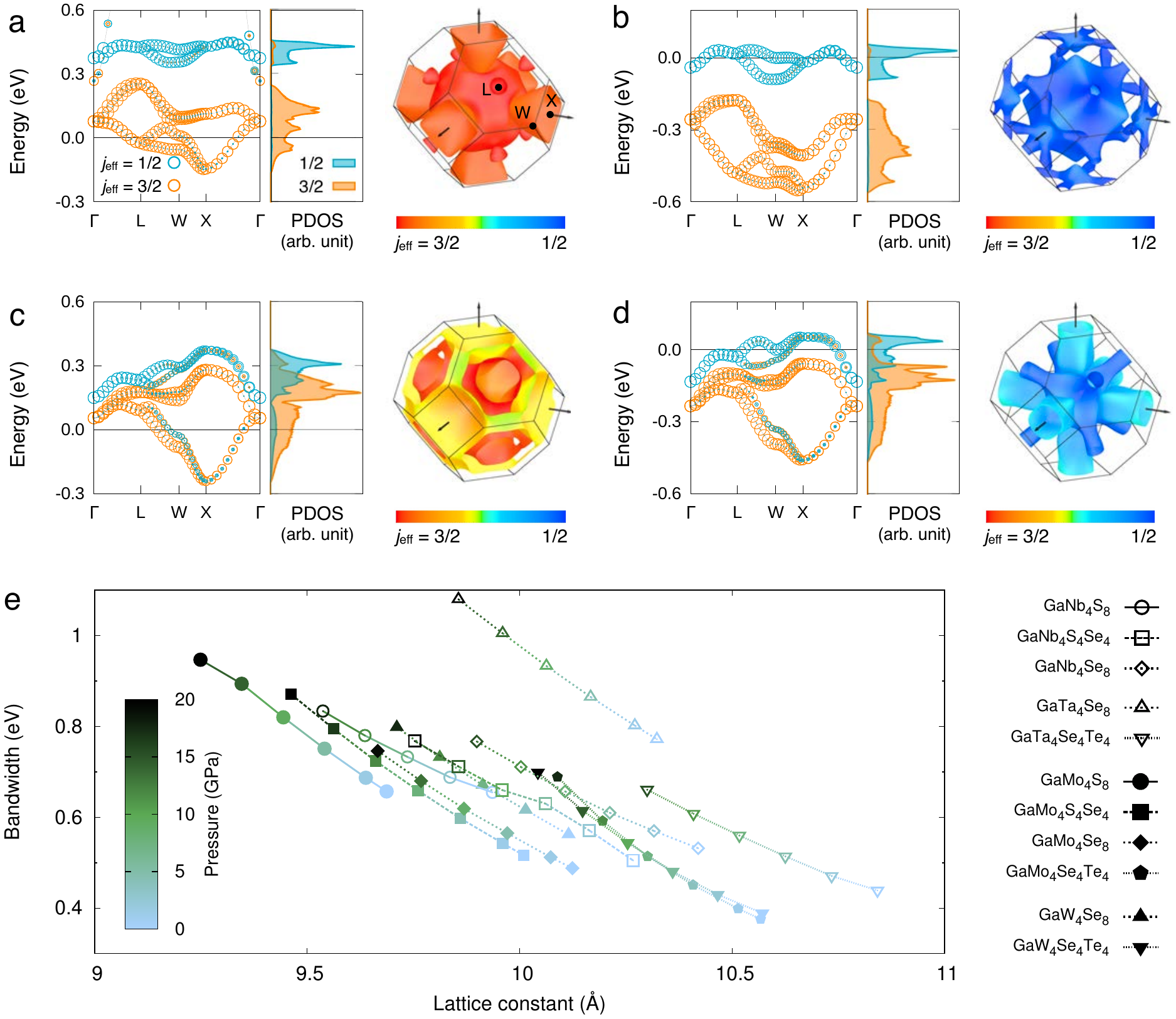}
  \caption{
        }
  \label{fig:APW}
\end{figure}

\textbf{Robust $\boldsymbol{j}_{\rm eff}$-ness in the Ga$M_4X_8$ series.}
The aforementioned $j_{\rm eff}$-ness in GaTa$_4$Se$_8$ remains robust in the Ga$M_4X_8$ series with a
neighboring 5$d$ transition metal ($M$ = W) as well as the 4$d$ counterparts ($M$ = Nb and Mo). Among the
series, $M$ = W compounds have not been reported previously in experiments. Thus we use optimized lattice
parameters by structural relaxations. In Fig.~\ref{fig:APW} {\bf a}-{\bf d}, the electronic structures of
GaTa$_4$Se$_4$Te$_4$~\cite{GTST}, GaW$_4$Se$_4$Te$_4$, GaNb$_4$Se$_8$~\cite{Pocha:JACS},
and GaMo$_4$Se$_8$~\cite{GMSe} are shown -- band structure, PDOS, and Fermi surface with projection onto
the molecular $j_{\rm eff}$ states. In Fig.~\ref{fig:APW} {\bf a} and {\bf b}, one can see the clear
separation and identification of the higher $j_{\rm eff}$ = 1/2 doublet and the lower $j_{\rm eff}$ = 3/2
quartet driven by the large SOC of the 5$d$ transition metal atoms. The overall band dispersions are
quite similar except for the location of the Fermi level; the $M$ = Ta and $M$ = W lacunar spinels are
well characterized by the quarter-filled $j_{\rm eff}$ = 3/2 and the half-filled $j_{\rm eff}$ = 1/2 systems,
respectively. In 4$d$ compounds, the separation between the $j_{\rm eff}$ subbands is reduced due to the
smaller SOC compared with the 5$d$ systems (Fig.~\ref{fig:APW} {\bf c} and {\bf d}). Nevertheless, there
is a discernible splitting between the $j_{\rm eff}$ = 1/2 and 3/2 bands, which is comparable to or even
better than that in the prototype $j_{\rm eff}$ compounds, Sr$_2$IrO$_4$ and Ba$_2$IrO$_4$~\cite{Arita}.

To acquire a well-identified $j_{\rm eff}$ band, we need the $j_{\rm eff}$ state as a local basis,
and the inter-orbital hopping terms between the $j_{\rm eff}$ subspaces should be suppressed.
Hence, there are three important conditions to realize the ideal $j_{\rm eff}$ system:
high symmetry protecting the $l_{\rm eff}$=1 three-fold orbital degeneracy, small
bandwidth minimizing the inter-orbital mixing, and large SOC fully entangling the spin and orbital degrees
of freedom. The lacunar spinel compounds comfortably satisfy the above conditions; the tetrahedral
symmetry of the $M_4$ cluster protects the orbital degeneracy, the long inter-cluster distance leads to
the small bandwidth, and a large SOC is inherent in 4$d$ and 5$d$ transition metal atoms.

Figure~\ref{fig:APW} {\bf e} introduces one important controlling parameter -- the bandwidth. By changing
the inter-cluster distance via external pressure and/or by substituting chalcogen atoms, the bandwidth of the
molecular $t_2$ band can be tuned over a wide range. In the $M$ = Ta series, for example, the bandwidth varies
from 0.4 to 1.1~eV. Consequently, the effective electron correlation strength, given by the ratio between
the bandwidth and the on-site Coulomb interactions, can be controlled to reach from the weak to the
strong coupling regime. In fact, the bandwidth-controlled insulator-to-metal transitions were observed in
GaTa$_4$Se$_4$ and GaNb$_4$Se$_4$~\cite{TaPhuoc,AbdElmeguid:PRL}, implying that both the weakly and strongly
interacting limits are accessible in a single compound.

\textbf{Effective Hamiltonian.}
From the apparent separation between the $j_{\rm eff}$ subbands, as well as the similar band dispersions,
the Ga$M_4X_8$ series are governed by a common effective Hamiltonian composed of two independent $j_{\rm eff}$
= 1/2 and 3/2 subspaces, {\it i.e.} $\mathcal{H}_{\rm eff} \simeq \mathcal{H}^{1/2} \oplus \mathcal{H}^{3/2}$.
(See Supplementary Note 2 and 3.) 
Therefore, the compounds with $M$ = Nb/Ta and $M$ = Mo/W are described by
the quarter-filled $\mathcal{H}^{3/2}$ and the half-filled $\mathcal{H}^{1/2}$ systems, respectively. The
nearest-neighbor hopping terms for each subspace are written as
\begin{eqnarray}
  \mathcal{H}^\tau_{\rm hopping} &=& \sum_{\langle ij \rangle}
  {\bf C}^\dag_{i\tau} {\bf T}^\tau_{ij} {\bf C}_{j\tau} ~~ \left( \tau=1/2,~3/2 \right), \\
  \mathrm{with} ~~{\bf T}^{1/2}_{ij} &=& t^0 {\bf I} + i\boldsymbol{t}^{\rm D}_{ij} \cdot {\bf S}^{1/2} \nonumber \\
  {\bf T}^{3/2}_{ij} &=& t^0 {\bf I} + i\boldsymbol{t}^{\rm D}_{ij} \cdot {\bf S}^{3/2}
  + \boldsymbol{t}^{\rm Q}_{ij} \cdot \mathbf{\Gamma}, \nonumber
  \label{eq:jeffhop}
\end{eqnarray}
where ${\bf S}^{1/2}$ and ${\bf S}^{3/2}$ are the $j_{\rm eff}$ = 1/2 and 3/2 pseudospin matrices, respectively,
and $\mathbf{\Gamma}$ are the 5-component Dirac Gamma matrices. $t^0$ and $\boldsymbol{t}^{\rm Q}$'s are even, and
$\boldsymbol{t}^{\rm D}$'s are odd functions under the spatial inversion; $\boldsymbol{t}^{\rm D}$'s are allowed
by the inversion asymmetry of the $M_4$ cluster. The pseudospin-dependent hopping terms
$\boldsymbol{t}^{\rm D}$ and $\boldsymbol{t}^{\rm Q}$ can be interpreted as the effective magnetic dipolar
and quadrupolar fields acting on the hopping electron, respectively.

\begin{figure}
  \centering
  \includegraphics[width=0.7\textwidth]{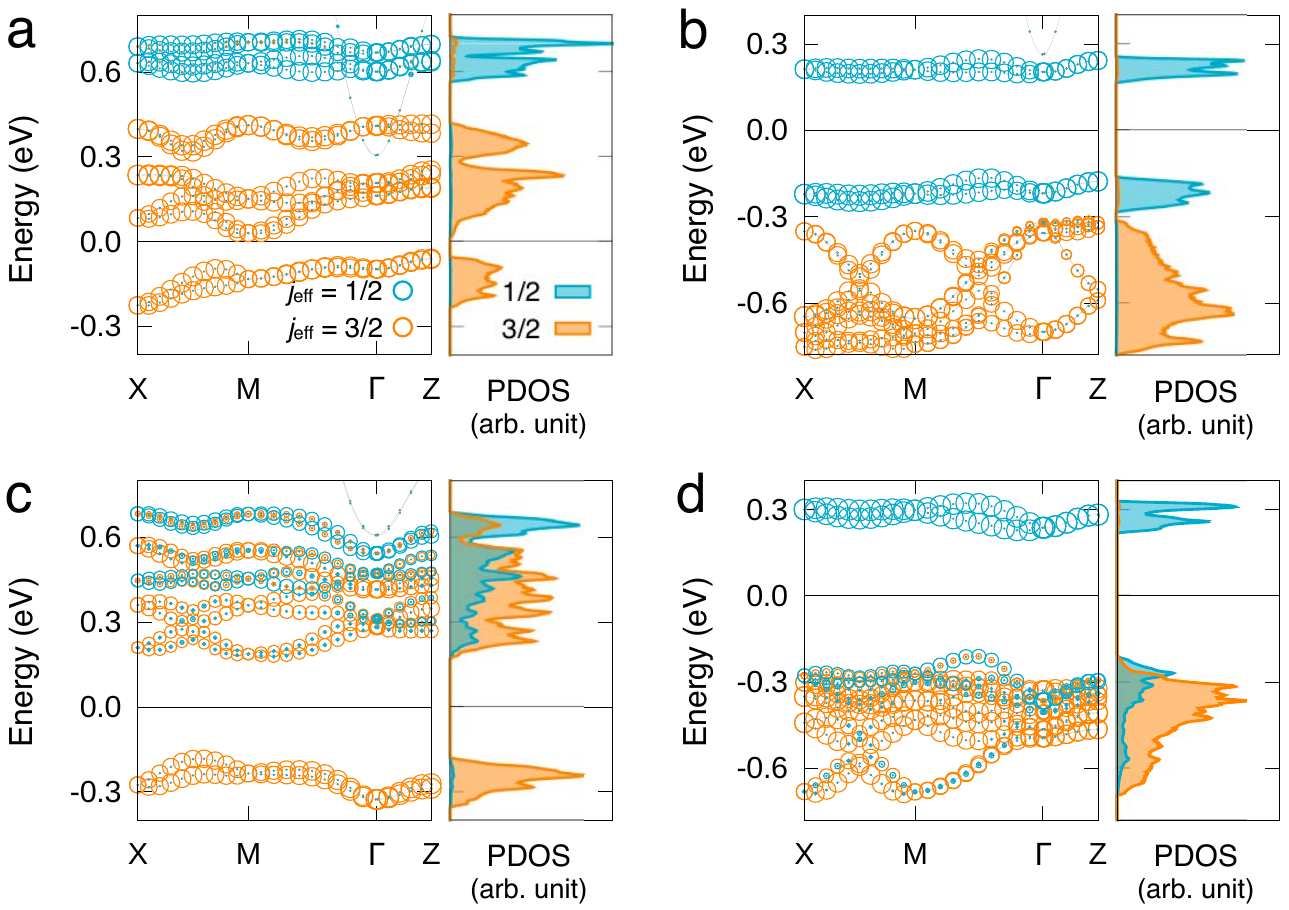}
  \caption{
  }
  \label{fig:DFT+U}
\end{figure}

\textbf{DFT+SOC+$\boldsymbol{U}$ calculations.}
So far, we have discussed about the $j_{\rm eff}$-ness without containing electron correlations, which provides
a valid picture in the weak coupling regime. Once taking electron correlations into account, one important
question arises on the robustness of the molecular $j_{\rm eff}$ states under the influence of the on-site Coulomb
interaction. To answer this question, we perform DFT+SOC+$U$ calculations for GaTa$_4$Se$_4$Te$_4$, GaW$_4$Se$_4$Te$_4$,
GaNb$_4$Se$_8$, and GaMo$_4$Se$_8$.
We consider two simplest magnetic configurations, ferromagnetic and antiferromagnetic order,
and the antiferromagnetic solutions for each compound are shown in Fig.~\ref{fig:DFT+U}.
In the 5$d$ compounds, the molecular $j_{\rm eff}$ states remain robust with developing a SOC-assisted Mott gap
within each $j_{\rm eff}$ subspace (Fig.~\ref{fig:DFT+U} {\bf a} and {\bf b}). For the 4$d$ compounds, the
$j_{\rm eff}$ character is enhanced from the non-interacting cases in Fig.~\ref{fig:APW} {\bf c} and {\bf d};
the occupied states in GaNb$_4$Se$_8$ (Fig.~\ref{fig:DFT+U} {\bf c}) and the unoccupied states in GaMo$_4$Se$_8$
(Fig.~\ref{fig:DFT+U} {\bf d}) are dominated by $j_{\rm eff}$=3/2 and 1/2 characters, respectively. The
strengthened $j_{\rm eff}$ character by the cooperation with electron correlations is consistent with the recent
theoretical results on Sr$_2$IrO$_4$~\cite{Arita,ZHV}. 
See the Supplementary Note 4, Supplementary Figure 3-6, and Supplementary Table 2-5 for more details.

\textbf{\large Discussion}

\begin{figure}
  \centering
  \includegraphics[width=1.05\textwidth]{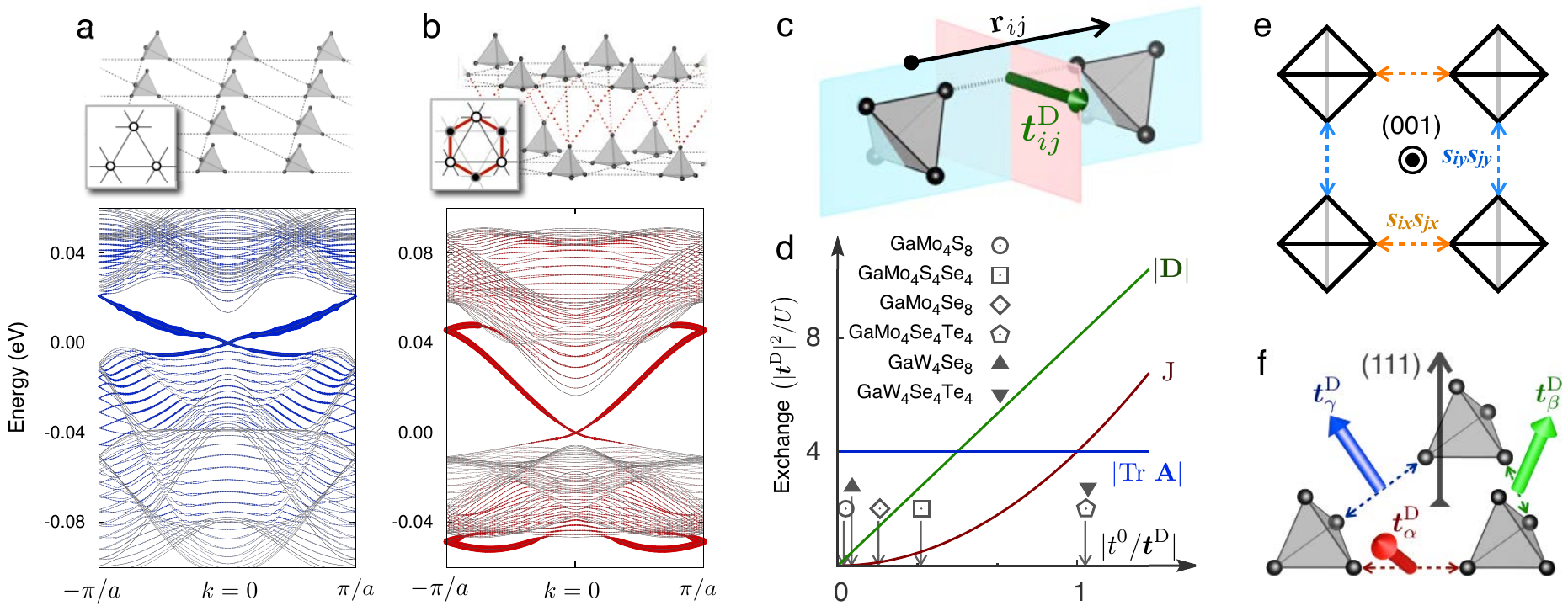}
  \caption{
 }
  \label{fig:edge}
\end{figure}

The effective Hamiltonian of the lacunar spinel series has intriguing implications both in the weak and strong
coupling regimes. As suggested in previous studies~\cite{Haldane,KM,Shitade:NIO}, the effective fields exerted
on the hopping electron can induce a topological insulating phase in the weak coupling regime.
In fact, a non-trivial band topology is realized within the molecular $j_{\rm eff}$ bands in thin film geometries:
the monolayer (Fig.~\ref{fig:edge} {\bf a}) and the bilayer thin film (Fig.~\ref{fig:edge} {\bf b}) of the $M_4$
clusters normal to the (111)-direction. Each system corresponds to the triangular and honeycomb lattice,
respectively, and the inter-layer coupling enhanced by a factor of three is adopted in the bilayer system.
Non-trivial gaps emerge in the half-filled $j_{\rm eff}$ = 3/2 bands in the monolayer and the half-filled
$j_{\rm eff}$ = 1/2 bands in the bilayer system.
A two-dimensional topological insulator phase is indicated by an odd number of edge Dirac cones
at time-reversal invariant momenta in ribbon geometries (Fig.~\ref{fig:edge} {\bf a} and {\bf b}).
Such two-dimensional geometries might be
feasible with the help of the state-of-the-art epitaxial technique prevailing in oxide perovskite compounds~\cite{DiXiao},
or by mechanically cleaving the single crystal to get clean surfaces
as done in previous studies on GaTa$_4$Se$_8$~\cite{Dubost,Dubost2}.

In the strong coupling regime, the large on-site Coulomb terms are added to the kinetic Hamiltonian, and the hopping
terms ${\bf T}^{\tau}_{ij}$ are treated as perturbations. The localized $j_{\rm eff}$ pseudospins become low-energy
degrees of freedom and exchange interactions between the neighboring $j_{\rm eff}$ moments emerge. In the simplest
example, the one-band Hubbard model within the half-filled $\mathcal{H}^{1/2}$, the resulting spin Hamiltonian
for the $j_{\rm eff}$ = 1/2 moments is written as~\cite{Micklitz,Jackeli}
\begin{equation}
  \mathcal{H}^{1/2}_{\rm spin} = \sum_{\langle ij \rangle} \left[
  {\rm J}{\bf s}_i\cdot{\bf s}_j +
  {\bf D}_{ij} \cdot \left( {\bf s}_i\times{\bf s}_j \right) +
  {\bf s}_i\cdot{\bf A}_{ij}\cdot{\bf s}_j
  \right],
  \label{eq:Hex}
\end{equation}
Among the exchange interaction terms, the Dzyaloshinskii-Moriya ${\bf D}_{ij}$ and the pseudodipolar interaction
${\bf A}_{ij}$ depend on $\boldsymbol{t}^{\rm D}_{ij}$, whose direction is determined by the two mirror planes,
as illustrated in Fig.~\ref{fig:edge} {\bf c} 
(details in Supplementary Note 5).
As shown in Fig.~\ref{fig:edge}
{\bf d}, the relative magnitude of each exchange term is changed with different chalcogen atoms, so that systematic
study of the anisotropic Hamiltonian in Eq.~\ref{eq:Hex} can be made in the $M$ = Mo/W compounds. Especially,
GaMo$_4$S$_8$ and GaW$_4$Se$_8$ satisfies the limit of $\vert t^0/ \boldsymbol{t}^{\rm D} \vert \rightarrow 0$,
where the spin Hamiltonian becomes highly anisotropic and bond-direction-dependent such that
\begin{align}
    \mathcal{H}^{1/2}_{\rm spin}
	&\rightarrow \sum_{\langle ij\rangle} {\bf s}_{i} \cdot {\bf A}_{ij} \cdot {\bf s}_{j} \nonumber \\
	&= \frac{4\vert \boldsymbol{t}^{\rm D} \vert^2}{U}\sum_{\langle ij \rangle}
	\left[ 2({\bf s}_i \cdot \hat{\boldsymbol{t}}^{\rm D}_{ij} ) ({\bf s}_j \cdot \hat{\boldsymbol{t}}^{\rm D}_{ij} )
    - {\bf s}_i \cdot {\bf s}_j \right],
\label{eq:genHCmodel}
\end{align}
with $\hat{\boldsymbol{t}}^{\rm D}_{ij} = \boldsymbol{t}^{\rm D}_{ij} / \vert \boldsymbol{t}^{\rm D} \vert$.
In addition to the Heisenberg term, the Hamiltonian contains the bond-dependent and Ising-like
pseudodipolar interaction, called as a Heisenberg-compass model~\cite{Review:compass}. It can be further
reduced to distinct two-dimensional spin models in thin film geometries. Figures~\ref{fig:edge} {\bf e}
and {\bf f} show two examples --- the (001)- and (111)-monolayer lead to the 90$^\circ$- and 60$^\circ$-compass
model with the Heisenberg exchange term on a square and a triangular lattice, respectively.

The $j_{\rm eff}$=3/2 systems in the strong coupling limit could also have a significant implication in terms
of unconventional multipolar orders~\cite{Jackeli:SVO,Chen:DP,Multipole}. On top of the nonmagnetic insulating behavior,
the weak tetragonal superstructure and the anomalous magnetic response observed in GaNb$_4$S$_8$ at
T$\sim$31~K~\cite{Jakob:GNS} could give some clues on the quadrupolar ordered phase as well as the spin liquid
phase suggested in Ref.~\onlinecite{Chen:DP}, which promptly calls for further research on the $j_{\rm eff}$=3/2
spin model.

The formation of the $M_4$ cluster and SOC are the essential requisites to realize the molecular $j_{\rm eff}$
state in these three-dimensional intermetallic compounds. The strong tetramerization sustains the isolated
molecular bands with three-fold orbital degeneracy and narrow bandwidth, and the large SOC fully entangles
the spin and orbital components. The existence of the pure quantum state has been shedding light on studying
the ideal quantum model systems in strongly correlated physics; the Hubbard Hamiltonian or the frustrated
spin Hamiltonian  based on the pure spin-half state have been realized in several organic compounds~\cite{
Yamashita,Kagawa,Yamashita04062010}. Likewise, the molecular form of the ideal $j_{\rm eff}$ state as a pure
quantum state might be of great use to explore the emergent phenomena in the spin-orbit coupled correlated
electron systems.

\textbf{\large Methods}

\textbf{First-principles calculations}~
Structural optimizations were done with the projector augmented wave potentials and the PBEsol~\cite{PBEsol}
generalized gradient approximation as implemented in the Vienna {\it ab-initio} Simulation Package~\cite{VASP1,VASP2}.
Momentum space integrations were performed on a 12$\times$12$\times$12 Monkhorst-Pack grid, and a 300~eV energy
cutoff was used for the plane-wave basis set. The force criterion was 10$^{-3}$~eV/\AA, and the pressures
exerted were estimated by using the Birch-Murnaghan fit.

For the electronic structure calculations, we used OPENMX code~\cite{OpenMX1} based on the
linear-combination-of-pseudo-atomic-orbital basis formalism. 400 Ry of energy cutoff was used for the real-space
integration. SOC was treated via a fully relativistic $j$-dependent pseudo potential in a non-collinear scheme.
Simplified DFT+$U$ formalism by Dudarev {\it et al.}~\cite{Dudarev}, implemented in OPENMX code~\cite{OpenMX:LDA+U},
was adopted in the DFT+SOC+$U$ calculations. $U_{\rm eff}\equiv U-J$ = 2.5 and 2.0~eV was used for the 4$d$ and 5$d$
compounds, respectively.

\textbf{Acknowledgments}~
We thank Yong-Baek Kim, Eun-Gook Moon, Tae-Won Noh, and Je-Geun Park for helpful discussions. This
work was supported by the Institute for Basic Science (IBS) in Korea. Computational resources were
provided by the National Institute of Supercomputing and Networking/Korea Institute of Science and
Technology Information with supercomputing resources including technical support (Grant No. KSC-2013-C2-005).

\textbf{Competing financial interests}~
The authors declare no competing financial interests.

\newpage
\newpage

\clearpage

\clearpage

\clearpage

\clearpage

{\bf Figure~1 $\vert$}
{\bf Molecular form of spin-orbital entangled $\boldsymbol{j}_{\rm eff}$ states in GaTa$_4$Se$_8$.}
{\bf a} The connectivity between the neighboring $M_4$ clusters and the local distortion of each cluster.
{\bf b} Band structure and projected density of states of GaTa$_4$Se$_8$ without SOC.
{\bf c} Three Wannier orbitals constructed from the triplet molecular orbital bands near the Fermi level.
{\bf d} Band structure and density of states with SOC, projected onto the $j_{\rm eff}=1/2$ and $3/2$
subspaces. The size of the circle in the band structure shows the weight of each subspace in each
Bloch state. \\

{\bf Figure~2 $\vert$}
{\bf $\boldsymbol{j}_{\rm eff}$-ness in the Ga{\emph M$_4$X$_8$} series.}
The molecular $j_{\rm eff}$-projected band structures, density of states, and the Fermi surfaces of
{\bf a} GaTa$_4$Se$_4$Te$_4$, {\bf b} GaW$_4$Se$_4$Te$_4$,
{\bf c} GaNb$_4$Se$_8$, and {\bf d} GaMo$_4$Se$_8$ are presented.
{\bf e} The relation between the external hydrostatic pressure,	 lattice constant, and bandwidth of the molecular
$t_2$ bands in the absence of SOC. \\

{\bf Figure~3 $\vert$}
{\bf DFT+SOC+$\boldsymbol{U}$ calculations.}
The $j_{\rm eff}$-projected band structure and density of states of
{\bf a} GaTa$_4$Se$_4$Te$_4$, {\bf b} GaW$_4$Se$_4$Te$_4$,
{\bf c} GaNb$_4$Se$_8$, and {\bf d} GaMo$_4$Se$_8$ with the presence of electron correlations
and antiferromagnetic order. \\

{\bf Figure~4 $\vert$}
{\bf Topological insulating phases and anisotropic spin model.}
The one-dimensional band structure of
{\bf a} half-filled $j_{\rm eff}$ = 3/2 monolayer and
{\bf b} half-filled $j_{\rm eff}$ = 1/2 bilayer $M_4$ ribbons (20 unit cell width).
The insets show schematic top view of each system, where the thin grey and the thick red lines represent
the intra- and the inter-planar bonding, respectively. The thickness of the colored fat lines in the band
structure represent the weights on the edge.
{\bf c} Two mirror planes (blue and red) existing in between the neighboring $M_4$ clusters
determine the direction of $\boldsymbol{t}^{\rm D}_{ij}$ illustrated as green arrow.
{\bf d} Magnitudes of Heisenberg (dark red), Dzyaloshinskii-Moriya (green), and pseudodipolar (blue)
exchange interactions as a function of $ \vert t_0 / \boldsymbol{t}^{\rm D} \vert$. The magnitude of
$ \vert t_0 / \boldsymbol{t}^{\rm D} \vert$ for each of the $M$ = Mo/W compounds is marked on the horizontal axis.
{\bf e} The 90$^\circ$- and {\bf f} the 60$^\circ$- compass interactions are
realized on (001) and (111) $M_4$ monolayers, respectively. \\

\newpage

\makeatletter
\renewcommand{\thefigure}{\@arabic\c@figure}
\makeatother

\makeatletter
\renewcommand\@biblabel[1]{}
\makeatother

\makeatletter
\renewcommand{\thetable}{\@arabic\c@table}
\makeatother

\makeatletter
\renewcommand{\theequation}{\@arabic\c@equation}
\makeatother

\newenvironment{correct}%
{\noindent\ignorespaces}%
{\par\noindent\ignorespacesafterend}

\setlength{\parindent}{0pt}

\textbf{\Large{Supplementary Figures}} \\

\begin{figure}[h]
  \centering
  \includegraphics[width=1.00\textwidth]{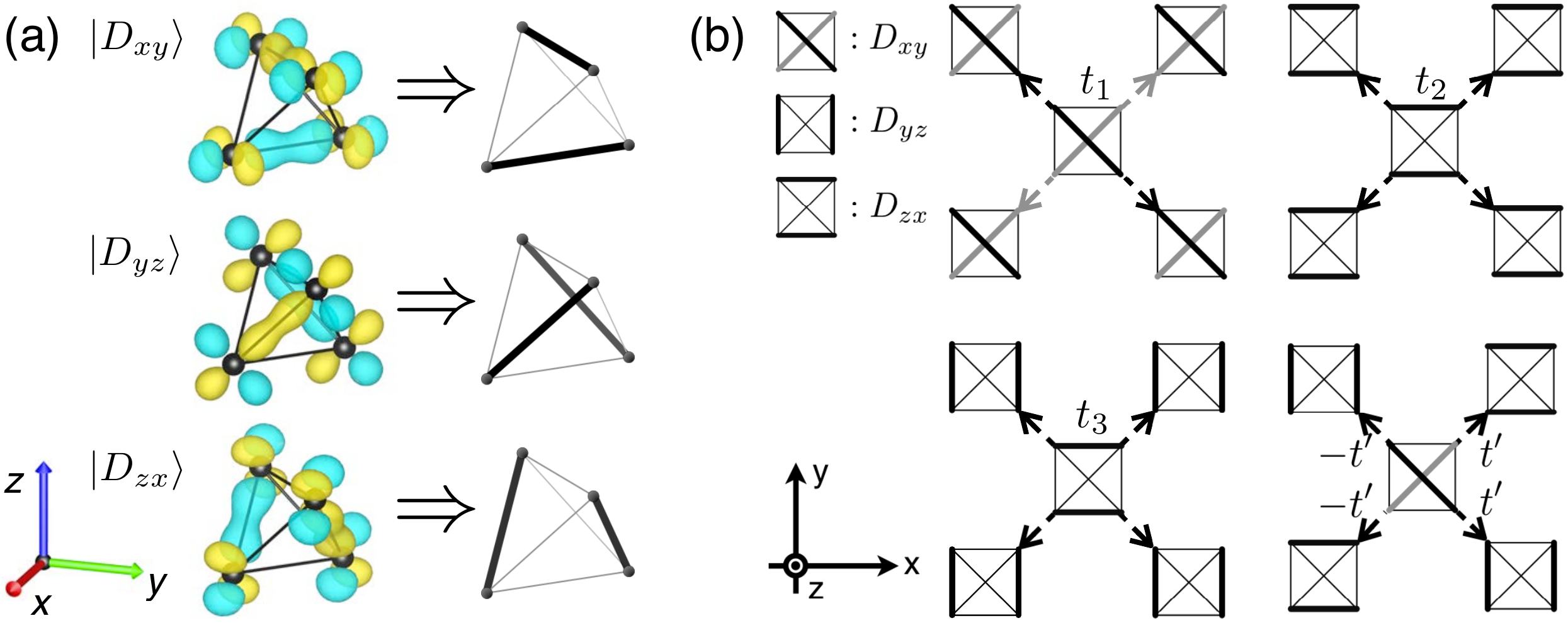}
  \label{figS:t2hop}
\end{figure}
{\bf Supplementary Figure 1: Molecular $\boldsymbol{t_2}$ orbitals and hopping channels.}
(a) The molecular $t_2$ orbitals and their schematic representations.
(b) Four nearest-neighbor hopping channels -- $t_1$, $t_2$, $t_3$, and $t'$ --
between the molecular $t_2$ orbitals on the $xy$-plane. \\

\begin{figure}[h]
  \centering
  \includegraphics[width=0.70\textwidth]{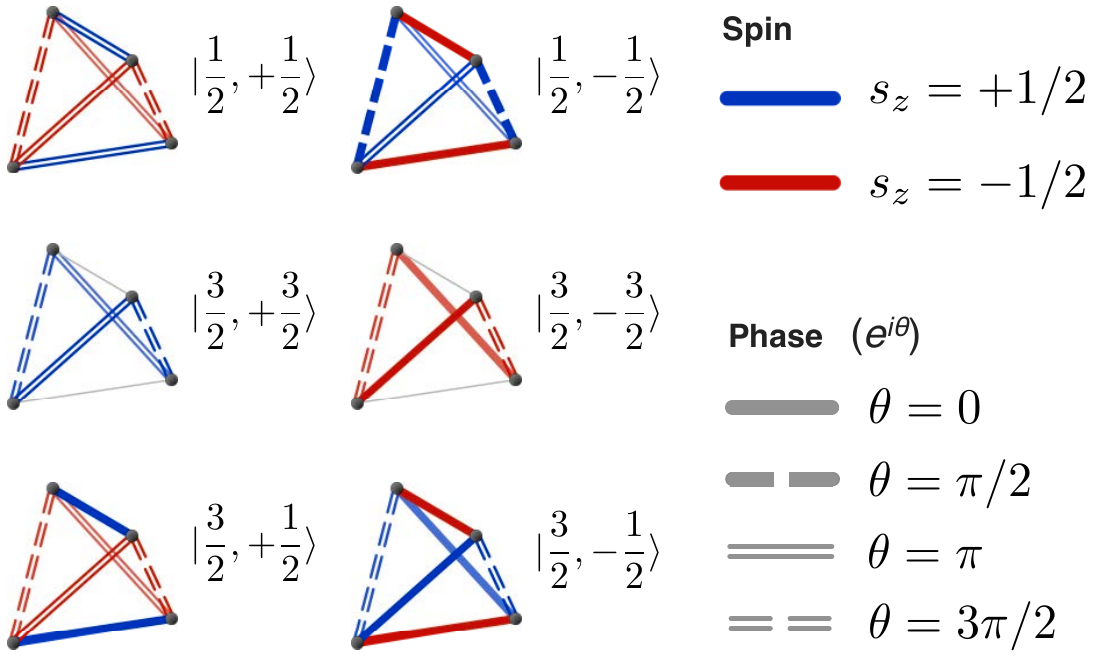}
  \label{figS:jeffstates}
\end{figure}
{\bf Supplementary Figure 2: Molecular $\boldsymbol{j}_{\rm eff}$ orbitals.}
Schematic viewgraph of the molecular $j_{\rm eff}$ orbitals.
Color and type of the thick lines represent the spin component and the phase factor
assigned to the constituent molecular $t_2$ orbitals. \\



\begin{figure}[h]
  \centering
  \includegraphics[width=0.8\textwidth]{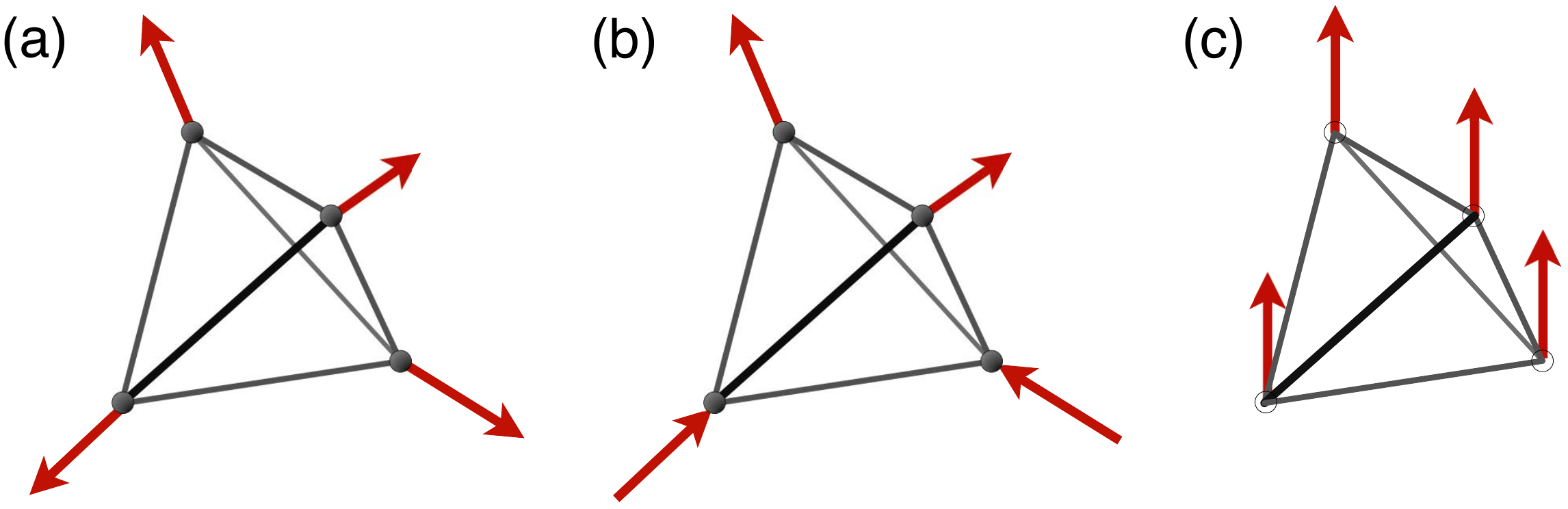}
  \label{figS:NCC}
\end{figure}
{\bf Supplementary Figure 3: Initial non-collinear magnetic configurations.}
Three initial magnetic configurations within the M$_4$ cluster used in the DFT+SOC+$U$ calculations:
(a) the all-in-all-out, (b) the 2-in-2-out, and (c) the collinear order. \\\\

\begin{figure}[h]
  \centering
  \includegraphics[width=0.9\textwidth]{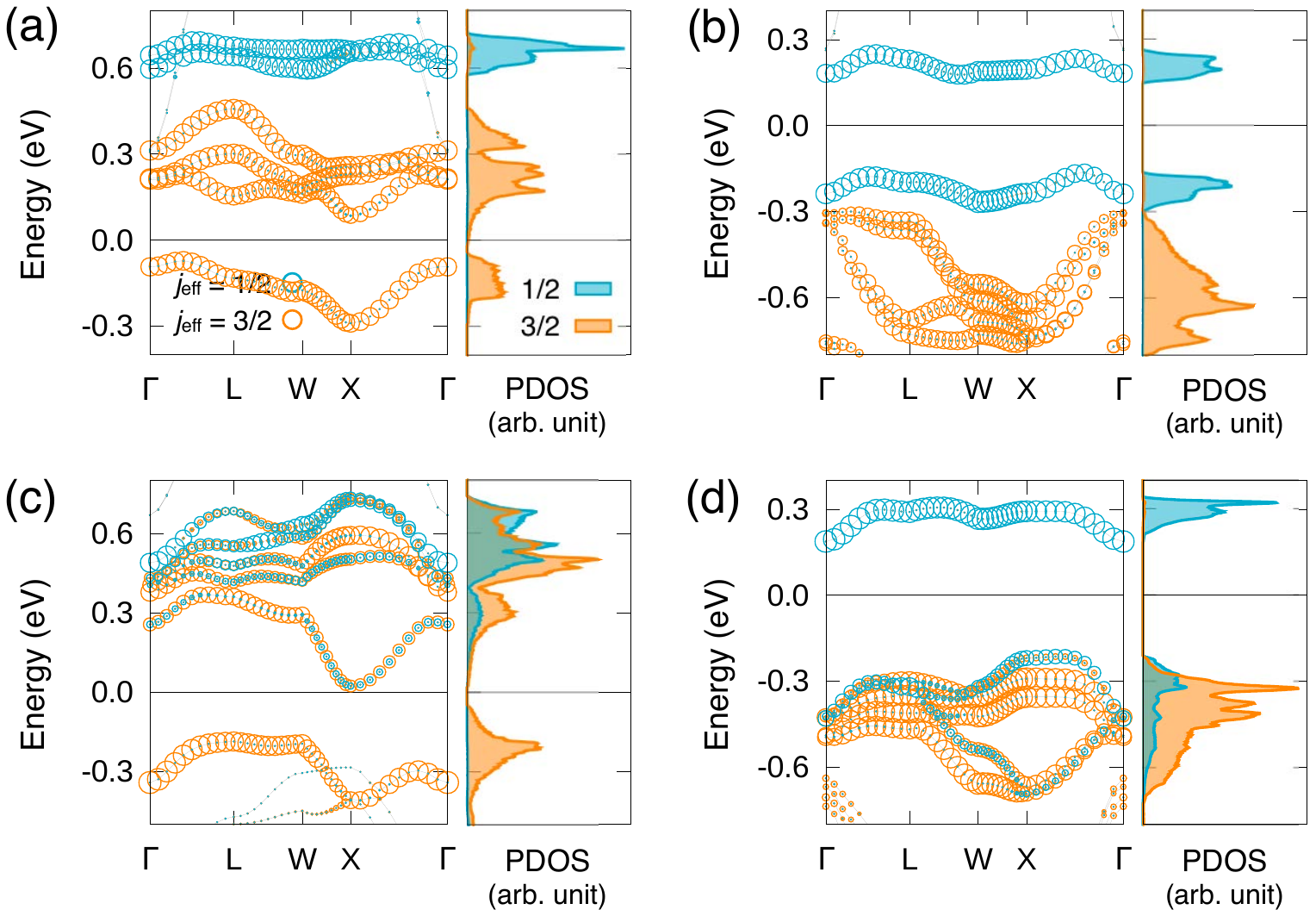}
  \label{figS:FMjeff}
\end{figure}
{\bf Supplementary Figure 4: $j_{\rm eff}$-projected electronic structures with ferromagnetic order.}
The $j_{\rm eff}$-projected band structures and PDOS of
(a) GaTa$_4$Se$_4$Te$_4$, (b) GaW$_4$Se$_4$Te$_4$,
(c) GaNb$_4$Se$_8$, and (d) GaMo$_4$Se$_8$ with the presence of electron correlations
and ferromagnetic order. \\

\begin{figure}[h]
  \centering
  \includegraphics[width=0.7\textwidth]{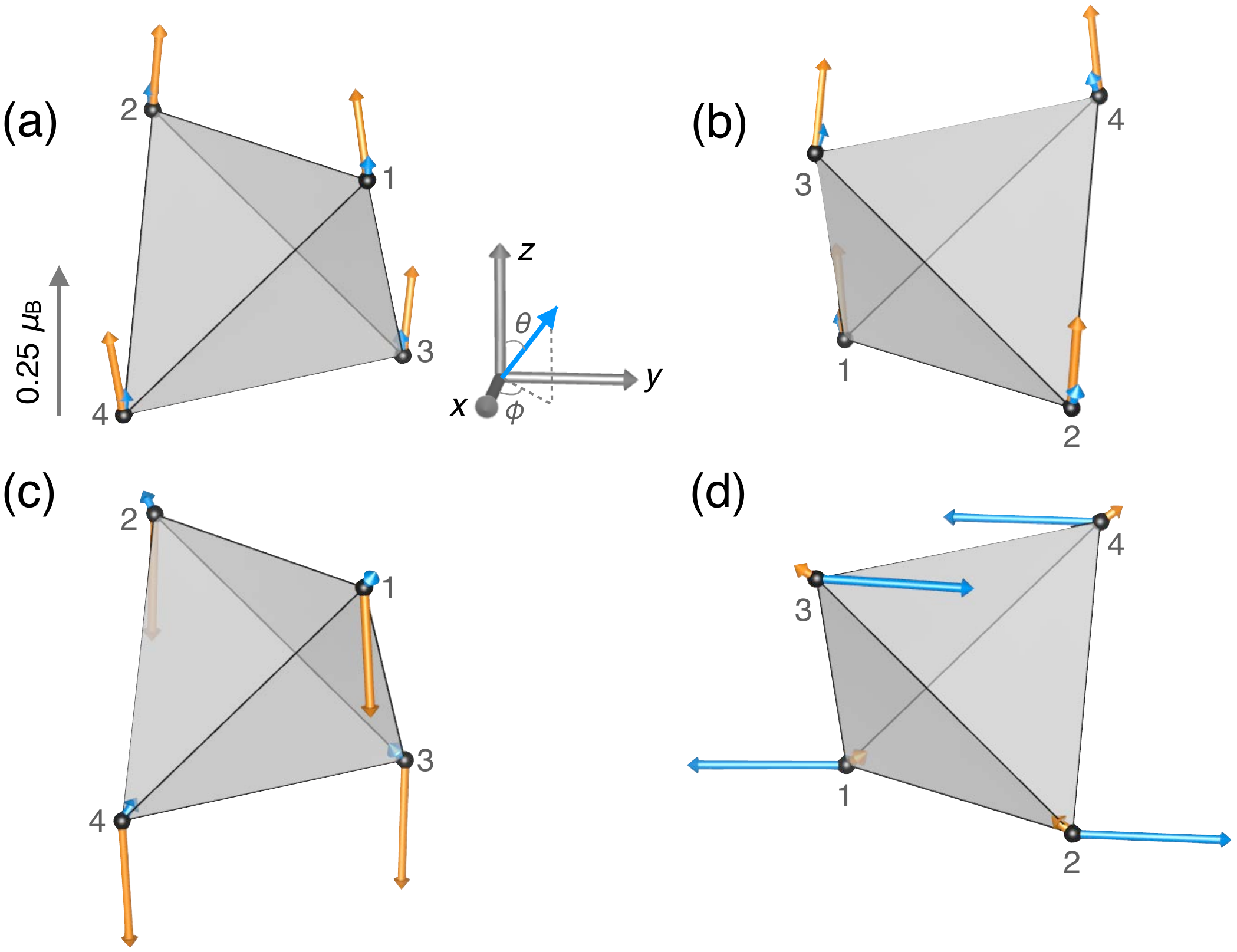}
  \label{figS:Mag}
\end{figure}
{\bf Supplementary Figure 5: Magnetic ordering in a $M_4$ cluster.}
The spin (blue) and orbital (orange) angular momenta at the $M_4$ corners of
(a) GaW$_4$Se$_4$Te$_4$, (b) GaMo$_4$Se$_8$,
(c) GaTa$_4$Se$_4$Te$_4$, and (d) GaNb$_4$Se$_8$ from the FM results (Supplementary Figure 4).
$M_4$ site indices for each compound ($i=1,\cdots,4$) are shown. \\\\

\begin{figure}[h]
  \centering
  \includegraphics[width=0.4\textwidth]{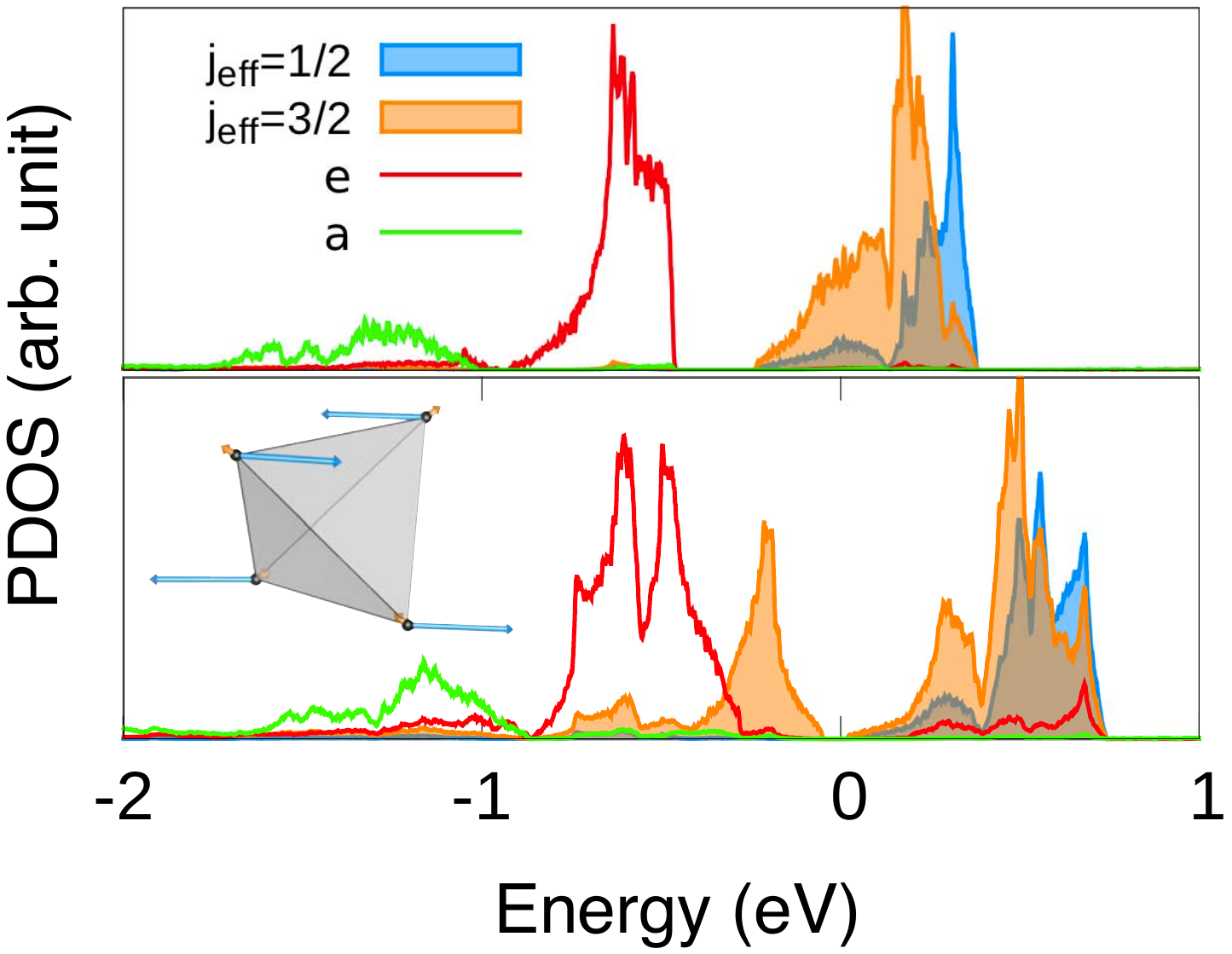}
  \label{figS:GNSe}
\end{figure}
{\bf Supplementary Figure 6: PDOS of GaNb$_4$Se$_8$ projected onto the molecular states.}
Top and bottom panel show PDOS from the paramagnetic result at $U_{\rm eff}$ = 0.0~eV
and the non-collinear order at $U_{\rm eff}$ = 2.5~eV, respectively.
The inset in the lower panel shows SAM (blue) and OAM (orange) within the $M_4$ cluster. \\

\clearpage

\textbf{\Large{Supplementary Tables}} \\

\begin{table}[h]
\centering
\begin{tabular}{ll|rrrrl} \hline\hline
                    & $P$ (GPa) & ~~~~~~~~~~ $t_1$ & ~~~~~~~~~~$t_2$ & ~~~~~~~~~~$t_3$ & ~~~~~~~~~~$t'$ & (meV)  \\ \hline
GaNb$_4$S$_8$      &~0  & -60.5 &  30.9 &  4.3 & 16.7 & \\
                   &21 & -86.3 &  44.7 &  11.0 & 18.7 & \\
GaNb$_4$S$_4$Se$_4$&~0  & -33.4 &  22.4 &  1.6  & 15.7 & \\
                   &20 & -50.6 &  32.9 &  4.3  & 17.4 & \\
GaNb$_4$Se$_8$     &~0  & -41.5 &  22.8 &  0.6  & 15.1 & \\
                   &15  & -89.3 & 48.5 & 15.4 & 23.4 & \\ \hline
GaMo$_4$S$_8$      &~0   & -48.0 & 24.9 & 7.4 & 19.9 & \\
                   &20  & -63.7 & 36.9 & 14.4 & 23.2 & \\
GaMo$_4$S$_4$Se$_4$&~0   & -22.9 & 20.4 & 3.1 & 18.1 & \\
                   &22  & -34.0 & 30.3 & 6.9 & 19.6 & \\
GaMo$_4$Se$_8$     &~0    & -31.0 & 19.7 & 2.8 & 16.7 & \\
                   &19   & -45.9 & 28.8 & 8.2 & 18.9 & \\
GaMo$_4$Se$_4$Te$_4$&~0    & -4.4  & 16.9 & 3.7 & 10.0 & \\
                   &17   & -3.6  & 26.6 & 5.0 & 8.2  & \\ \hline
GaTa$_4$Se$_8$     &~0    & -55.7 &  27.6 & 7.1 & 14.5 & \\
                   &20  & -75.5 &  37.1 & 8.4 & 15.1 & \\
GaTa$_4$Se$_4$Te$_4$&~0   & -22.9 & 17.1 & 12.0 & 9.2 & \\
                   & 15 & -33.8 & 24.5 & 15.7 & 8.2 & \\ \hline
GaW$_4$Se$_8$       &~0    & -42.3 &  22.6 & 7.3  & 16.7 & \\
                    &21    & -64.6 &  33.0 & 3.0 & 18.9 & \\
GaW$_4$Se$_4$Te$_4$ &~0     & -14.6 &  18.3 & 11.0 & 9.7 & \\
                    &20    & -20.6 &  23.6 & 16.6 & 8.9 & \\ \hline\hline
\end{tabular}
\label{tab:t2hop}
\end{table}
{\bf Supplementary Table 1: Molecular $\boldsymbol{t}_2$ hopping terms.}
NN hopping terms between the molecular $t_2$ orbitals of the lacunar spinel
compounds, with/without the external pressure. \\

\clearpage

\begin{table*}[h]
\centering
\footnotesize
\begin{tabular*}{\textwidth}{@{\extracolsep{\fill}}llrrrrrrrrrrrrrr} \hline\hline
$U_{\rm eff}$ &  &  \multicolumn{6}{c}{FM} &  & \multicolumn{6}{c}{AFM} & $\Delta E$ \\ \cline{3-8}\cline{10-15}
(eV) &$i$&
  $\vert S \vert$ & $\theta_S$ & $\phi_S$ & $\vert L \vert$ & $\theta_L$ & $\phi_L$ & &
  $\vert S \vert$ & $\theta_S$ & $\phi_S$ & $\vert L \vert$ & $\theta_L$ & $\phi_L$ & (meV) \\ \hline \noalign{\smallskip}
\multirow{4}{*}{1.0}
    &1& 0.041 & 2.5 & 225.0 & 0.125 & 8.3 & 225.0 &  & 0.032 & 30.4 & 44.9 & 0.103 & 29.7 & 44.9  & \multirow{4}{*}{2.9} \\
    &2& 0.041 & 6.2 & 45.0 & 0.127 & 11.5 & 45.0 &  & 0.042 & 35.2 & 44.9 & 0.139 & 35.3 & 44.9   & \\
    &3& 0.041 & 4.8 & 109.6 & 0.126 & 10.2 & 122.6 &  & 0.040 & 37.8 & 55.6 & 0.129 & 38.1 & 55.6 & \\
    &4& 0.041 & 4.8 & -19.6 & 0.126 & 10.2 & -32.6 &  & 0.040 & 37.8 & 34.3 & 0.129 & 38.1 & 34.3  & \\ \noalign{\medskip}
\multirow{4}{*}{1.5}
    &1& 0.044 & 5.5 & 45.0 & 0.139 & 4.9 & 225.0 &  & 0.038 & 33.6 & 46.4 & 0.117 & 29.7 & 46.9  & \multirow{4}{*}{-0.3} \\
    &2& 0.045 & 6.7 & 45.0 & 0.149 & 14.9 & 45.0 &  & 0.044 & 35.4 & 46.2 & 0.158 & 35.4 & 46.2  & \\
    &3& 0.044 & 6.2 & 50.6 & 0.144 & 12.1 & 100.8 &  & 0.043 & 36.5 & 51.9 & 0.148 & 38.4 & 56.9 & \\
    &4& 0.044 & 6.2 & 39.4 & 0.144 & 12.1 & -10.8 &  & 0.043 & 36.4 & 40.3 & 0.147 & 38.3 & 35.0 & \\ \noalign{\medskip}
\multirow{4}{*}{2.0}
    &1& 0.047 & 3.7 & 45.0 & 0.160 & 11.4 & 225.0 &  & 0.046 & 37.8 & 47.0 & 0.127 & 29.9 & 48.5  & \multirow{4}{*}{-0.01} \\
    &2& 0.047 & 6.7 & 225.0 & 0.158 & 9.0 & 45.0 &  & 0.043 & 35.9 & 47.2 & 0.174 & 35.6 & 47.2   & \\
    &3& 0.047 & 5.4 & -59.7 & 0.159 & 10.3 & 144.1 &  & 0.044 & 34.6 & 45.0 & 0.163 & 38.8 & 57.9 & \\
    &4& 0.047 & 5.4 & 149.7 & 0.159 & 10.3 & -54.1 &  & 0.044 & 34.6 & 49.7 & 0.162 & 38.7 & 35.6 & \\ \noalign{\smallskip}
\hline\hline
\end{tabular*}
\label{tab:W}
\end{table*}
{\bf Supplementary Table 2: Magnetism in GaW$_4$Se$_4$Te$_4$.}
Sizes (in $\mu_{\rm B}$) and directions (in degree) of SAM and OAM on the four corners of the $M_4$ cluster
for the FM and AFM configurations, the energy difference $\Delta E \equiv E_{\rm AFM} - E_{\rm FM}$,
and their $U_{\rm eff}$ dependence in GaW$_4$Se$_4$Te$_4$.
See Supplementary Figure 5 for the definitions of angle $\theta$, $\phi$, and site index $i$. \\

\begin{table*}[h]
\centering
\footnotesize
\begin{tabular*}{\textwidth}{@{\extracolsep{\fill}}llrrrrrrrrrrrrrr} \hline\hline
$U_{\rm eff}$ &  &  \multicolumn{6}{c}{FM} &  & \multicolumn{6}{c}{AFM} & $\Delta E$ \\ \cline{3-8}\cline{10-15}
(eV) &$i$&
  $\vert S \vert$ & $\theta_S$ & $\phi_S$ & $\vert L \vert$ & $\theta_L$ & $\phi_L$ & &
  $\vert S \vert$ & $\theta_S$ & $\phi_S$ & $\vert L \vert$ & $\theta_L$ & $\phi_L$ & (meV) \\ \hline \noalign{\smallskip}
\multirow{4}{*}{1.5}
    &1& 0.048 & 24.4 & 225.0 & 0.129 & 7.0 & 225.0 &  & 0.016 & 33.3 & 225.3 & 0.114 & 31.0 & 44.9	  & \multirow{4}{*}{-0.2} \\
    &2& 0.048 & 24.6 & 45.0 & 0.129 & 7.2 & 45.0 &  & 0.056 & 34.8 & 44.9 & 0.137 & 35.4 & 44.9	  & \\
    &3& 0.048 & 24.5 & 134.6 & 0.129 & 7.1 & 133.9 &  & 0.054 & 50.1 & 80.5 & 0.129 & 37.6 & 51.4	  & \\
    &4& 0.048 & 24.5 & -44.6 & 0.129 & 7.1 & -43.9 &  & 0.054 & 50.1 & 9.5 & 0.129 & 37.6 & 38.4     & \\ \noalign{\medskip}
\multirow{4}{*}{2.0}
    &1& 0.049 & 24.7 & 224.8 & 0.146 & 6.3 & 224.4 &  & 0.017 & 38.5 & 226.9 & 0.127 & 31.1 & 44.3 & \multirow{4}{*}{-0.1} \\
    &2& 0.050 & 25.4 & 45.2 & 0.147 & 7.9 & 45.5 &  & 0.058 & 35.1 & 44.4 & 0.154 & 35.5 & 44.5    & \\
    &3& 0.050 & 25.1 & 132.3 & 0.147 & 7.2 & 126.8 &  & 0.057 & 50.6 & 81.0 & 0.146 & 37.6 & 51.5  & \\
    &4& 0.050 & 25.0 & -42.3 & 0.147 & 7.1 & -36.7 &  & 0.057 & 50.4 & 8.8 & 0.146 & 37.7 & 37.6   & \\ \noalign{\medskip}
\multirow{4}{*}{2.5}
    &1& 0.051 & 24.0 & 225.0 & 0.164 & 7.1 & 225.0 &  & 0.017 & 42.3 & 226.8 & 0.140 & 31.3 & 44.3	 & \multirow{4}{*}{-0.1} \\
    &2& 0.051 & 23.9 & 45.0 & 0.164 & 7.0 & 45.0 &  & 0.059 & 35.4 & 44.5 & 0.170 & 35.6 & 44.5	 & \\
    &3& 0.051 & 24.0 & 135.2 & 0.164 & 7.1 & 135.5 &  & 0.059 & 50.3 & 81.5 & 0.161 & 37.8 & 51.7	 & \\
    &4& 0.051 & 24.0 & -45.2 & 0.164 & 7.1 & -45.5 &  & 0.060 & 50.2 & 8.3 & 0.161 & 37.9 & 37.4    & \\ \noalign{\smallskip}
\hline\hline
\end{tabular*}
\label{tab:Mo}
\end{table*}
{\bf Supplementary Table 3: Magnetism in GaMo$_4$Se$_8$.} \\

\clearpage

\begin{table*}[h]
\centering
\footnotesize
\begin{tabular*}{\textwidth}{@{\extracolsep{\fill}}llrrrrrrrrrrrrrr} \hline\hline
$U_{\rm eff}$ &  &  \multicolumn{6}{c}{FM} &  & \multicolumn{6}{c}{AFM} & $\Delta E$ \\ \cline{3-8}\cline{10-15}
(eV) &$i$&
  $\vert S \vert$ & $\theta_S$ & $\phi_S$ & $\vert L \vert$ & $\theta_L$ & $\phi_L$ & &
  $\vert S \vert$ & $\theta_S$ & $\phi_S$ & $\vert L \vert$ & $\theta_L$ & $\phi_L$ & (meV) \\ \hline \noalign{\smallskip}
1.0 && \multicolumn{13}{c}{Paramagnetic} & --- \\ \noalign{\medskip}
\multirow{4}{*}{1.5}
    &1& 0.033 & 27.9 & 45.0 & 0.165 & 175.9 & 45.0 &  & 0.038 & 35.6 & 225.0 & 0.051 & 149.6 & 45.0 & \multirow{4}{*}{-1.6} \\
    &2& 0.034 & 29.2 & 226.0 & 0.165 & 176.8 & 225.0 &  & 0.028 & 37.8 & 225.0 & 0.036 & 158.8 & 45.0 & \\
    &3& 0.034 & 28.8 & -46.7 & 0.165 & 176.3 & -37.9 &  & 0.036 & 38.9 & 210.2 & 0.055 & 140.0 & 38.4 & \\
    &4& 0.033 & 28.3 & 135.7 & 0.165 & 176.3 & 127.9 &  & 0.036 & 38.9 & 239.8 & 0.055 & 140.0 & 51.6 & \\ \noalign{\medskip}
\multirow{4}{*}{2.0}
    &1& 0.049 & 43.1 & 45.0 & 0.229 & 176.9 & 45.0 &  & 0.140 & 21.1 & 225.0 & 0.226 & 156.7 & 45.0 & \multirow{4}{*}{6.5} \\
    &2& 0.049 & 42.8 & 225.0 & 0.229 & 176.9 & 225.0 &  & 0.121 & 28.3 & 225.0 & 0.222 & 167.6 & 45.0 & \\
    &3& 0.049 & 43.0 & -44.8 & 0.229 & 176.9 & -45.0 &  & 0.112 & 2.4 & 0.3 & 0.221 & 162.8 & 24.0 & \\
    &4& 0.049 & 43.0 & 134.8 & 0.229 & 176.9 & 135.0 &  & 0.112 & 2.4 & 89.8 & 0.221 & 162.8 & 66.0 & \\ \noalign{\smallskip}
\hline\hline
\end{tabular*}
\label{tab:Ta}
\end{table*}
{\bf Supplementary Table 4: Magnetism in GaTa$_4$Se$_4$Te$_4$.} \\

\begin{table*}[h]
\centering
\footnotesize
\begin{tabular*}{\textwidth}{@{\extracolsep{\fill}}llrrrrrrrrrrrrrr} \hline\hline
$U_{\rm eff}$ &  &  \multicolumn{6}{c}{FM} &  & \multicolumn{6}{c}{AFM} & $\Delta E$ \\ \cline{3-8}\cline{10-15}
(eV) &$i$&
  $\vert S \vert$ & $\theta_S$ & $\phi_S$ & $\vert L \vert$ & $\theta_L$ & $\phi_L$ & &
  $\vert S \vert$ & $\theta_S$ & $\phi_S$ & $\vert L \vert$ & $\theta_L$ & $\phi_L$ & (meV) \\ \hline \noalign{\smallskip}
1.5 && \multicolumn{13}{c}{Paramagnetic} & --- \\ \noalign{\medskip}
\multirow{4}{*}{2.0}
    &1& 0.190 & 125.3 & 225.0 & 0.119 & 16.2 & 225.0 &  & 0.196 & 58.9 & 225.0 & 0.029 & 161.8 & 45.0 & \multirow{4}{*}{-17.4} \\
    &2& 0.112 & 140.5 & 225.0 & 0.128 & 18.9 & 45.0 &  & 0.137 & 82.1 & 45.0 & 0.104 & 141.9 & 225.0  & \\
    &3& 0.112 & 116.8 & 239.7 & 0.124 & 17.9 & 125.4 &  & 0.094 & 83.1 & 7.0 & 0.078 & 130.6 & 235.9  & \\
    &4& 0.112 & 116.8 & 210.3 & 0.124 & 17.9 & -35.4 &  & 0.094 & 83.1 & 83.0 & 0.078 & 130.6 & 214.1 & \\ \noalign{\medskip}
\multirow{4}{*}{2.5}
    &1& 0.269 & 90.5 & -87.0 & 0.047 & 56.0 & 72.2 &  & 0.294 & 110.2 & -87.9 & 0.061 & 56.1 & 70.3   & \multirow{4}{*}{-10.6} \\
    &2& 0.269 & 91.6 & 93.1 & 0.048 & 55.7 & 252.0 &  & 0.328 & 45.0 & 90.0 & 0.011 & 135.0 & 270.0   & \\
    &3& 0.269 & 91.8 & 86.9 & 0.048 & 55.6 & -72.1 &  & 0.175 & 45.0 & 90.0 & 0.003 & 135.0 & -90.0   & \\
    &4& 0.268 & 90.5 & 267.0 & 0.047 & 56.0 & 107.9 &  & 0.294 & 159.7 & 264.2 & 0.061 & 38.6 & 116.7 & \\ \noalign{\smallskip}
\hline\hline
\end{tabular*}
\label{tab:Nb}
\end{table*}
{\bf Supplementary Table 5: Magnetism in GaNb$_4$Se$_8$.}

\clearpage

\textbf{\Large{Supplementary Notes}} \\

\setlength{\parindent}{20pt}

\noindent
\textbf{Supplementary Note 1: Molecular $\boldsymbol{t}_2$ Hopping Terms.~}
In this section, we construct the tight-binding Hamiltonian based on the molecular $t_2$ states
in the absence of spin-orbit coupling (SOC).
Regarding only the nearest-neighboring(NN) sites $i$ and $j$,
the tight-binding Hamiltonian can be written as
\begin{equation}
\mathcal{H}^{t_2}_{{\rm hopping};ij} =
\left(
\begin{array}{ccc}
 S_1 & S_5 - A_2 & S_4 + A_1   \\
 S_5 + A_2 & S_2 & S_6 - A_3   \\
 S_4 - A_1 & S_6 + A_3 & S_3
\end{array}
\right)
\label{eq:H12}
\end{equation}
in terms of the basis set
$\left(
D_{xy},
D_{yz},
D_{zx}
\right)$.
Here, $S$ and $A$ denote symmetric and antisymmetric hopping terms with respect to the site inversion
$i \leftrightarrow j$ such that $\mathcal{H}^{t_2}_{{\rm hopping};ji} = \mathcal{H}^{t_2}_{{\rm hopping};ij} (-A)$,
or equivalently $\left(\mathcal{H}^{t_2}_{{\rm hopping};ij}\right)^T = \mathcal{H}^{t_2}_{{\rm hopping};ji}$.

Wannier function analysis shows that only 4 NN hopping channels -- say, $t_1$, $t_2$, $t_3$, and $t'$ --
are allowed in the $AM_4X_8$ compounds.
The edge-sharing geometry of the distorted $MX_8$ octahedra enables the correspondence of our $t_1$, $t_2$,
and $t_3$ hopping terms to those in the layered iridates $A_2$IrO$_3$ ($A$ = Li, Na)$^{1,2}$
as shown schematically in Supplementary Figure 1; $t_1$, $t_2$, and $t_3$ correspond to $t_{dd1}$ ($\sigma$-type),
$t_{pd}$ ($\pi$-type), and $t_{dd2}$ ($\delta$-type) hopping integrals in Supplementary Reference 1, respectively.
The antisymmetric term $t'$ is allowed due to the lack of inversion symmetry by the formation of the $M_4$ clusters.
Along the direction to the 12 NNs in the face-centered cubic lattice, {\it i.e.}
${\bf r}_{ij} = n_1 {\bf a}_1 + n_2 {\bf a}_2 + n_1 {\bf a}_2$, the NN hopping terms are as follows:\\
{\small
\begin{correct}
\phantom{AAAAA}($n_1$,$n_2$,$n_3$)=({\bf $\pm$1,0,0})
~~$S_1=t_1, ~ S_2=S_3=t_2, ~ S_6=-t_3, ~ A_1=-A_2 = \mp t'$\\
\phantom{AAAAA}($n_1$,$n_2$,$n_3$)=({\bf 0,$\pm$1,0})
~~$S_1=S_3=t_2, ~ S_2=t_1, ~ S_4=-t_3, ~ A_2=-A_3 = \mp t'$\\
\phantom{AAAAA}($n_1$,$n_2$,$n_3$)=({\bf 0,0,$\pm$1})
~~$S_1=S_2=t_2, ~ S_3=t_1, ~ S_5=-t_3, ~ A_1= -A_3 = \pm t'$\\
\phantom{AAAAA}($n_1$,$n_2$,$n_3$)=({\bf $\pm$1,$\mp$1,0})
~~$S_1=S_2=t_2, ~ S_3=t_1, ~ S_5=t_3, ~ A_1=A_3 = \pm t'$\\
\phantom{AAAAA}($n_1$,$n_2$,$n_3$)=({\bf 0,$\pm$1$,\mp$1})
~~$S_1=t_1, ~ S_2=S_3=t_2, ~ S_6=t_3, ~ A_1=A_2 = \pm t'$\\
\phantom{AAAAA}($n_1$,$n_2$,$n_3$)=({\bf $\pm$1,0,$\mp$1})
~~$S_1=S_3=t_2, ~ S_2=t_1, ~ S_4=t_3, ~ A_2=A_3 = \mp t'$
\end{correct}
}
Here we adopt the convention that $t_3,t'>0$. The other terms not shown above are all zero.
The amount of each hopping term, with and without external pressure, is shown in Supplementary Table 1.
The values of the NN hopping terms for GaTa$_4$Se$_8$, in the absence of external pressure,
are consistent with the previous work$^{3}$. \\

\noindent
\textbf{Supplementary Note 2: Molecular $\boldsymbol{j}_{\rm eff}$ Hopping Terms in the Presence of SOC.~}
As mentioned in the main text, the molecular $t_2$ states behave in the same way as the atomic $t_{\rm 2g}$
states do under SOC$^{4}$. The SOC Hamiltonian is written as
\begin{equation}
\mathcal{H}_{\rm SO} \equiv \lambda_{\rm SO} {\bf L} \cdot {\bf S}.
\label{eq:Hsoc}
\end{equation}
where $\lambda_{\rm SO}$ is the SOC strength of the transition metal atoms,
and ${\bf L}$ and ${\bf S}$ are the orbital and the spin angular momentum operators, respectively.

The eigenstates of the SOC Hamiltonian in Eq.~(\ref{eq:Hsoc}) are written as
\begin{eqnarray}
\vert j_{\rm eff}=\frac{1}{2};\pm \frac{1}{2} \rangle
&=& \mp \frac{1}{\sqrt{3}} \left(
   \vert D_{xy},\uparrow\downarrow \rangle
\pm\vert D_{yz},\downarrow\uparrow \rangle
+i \vert D_{xz},\downarrow\uparrow \rangle
\right) \nonumber \\
\vert j_{\rm eff}=\frac{3}{2};\pm \frac{1}{2} \rangle
&=& \sqrt{\frac{2}{3}} \left[
\vert D_{xy}, \uparrow\downarrow \rangle \mp
\frac{
\vert D_{yz}, \downarrow\uparrow \rangle \pm i \vert D_{zx},\downarrow\uparrow \rangle
}{2}
\right] \nonumber \\
\vert j_{\rm eff}=\frac{3}{2};\pm \frac{3}{2} \rangle
&=& \mp \frac{1}{\sqrt{2}} \left(
\vert D_{yz}, \uparrow\downarrow \rangle \pm
i \vert D_{zx}, \uparrow\downarrow \rangle
\right),
\label{eq:jbasis}
\end{eqnarray}
which are schematically shown in Supplementary Figure 2.
By adding SOC to the molecular $t_2$ Hamiltonian in Eq.~(\ref{eq:H12}) and transforming the molecular
$t_2$ into the $j_{\rm eff}$ basis sets, Hamiltonian now has the following form:
\begin{equation}
\mathcal{H}^{j_{\rm eff}} = \left(
\begin{array}{cc|cc}
   +\lambda_{\rm SO}  {\bf I}^{1/2} & & \mathbf{T}^{1/2}_{ij} & \boldsymbol{\Theta}_{ij}   \\
 & -\frac{1}{2}\lambda_{\rm SO} {\bf I}^{3/2}   &\boldsymbol{\Theta}_{ij}(-A)^\dag & \mathbf{T}^{3/2}_{ij}   \\ \hline
   \mathbf{T}^{1/2}_{ji} & \boldsymbol{\Theta}_{ji} &   +\lambda_{\rm SO} {\bf I}^{1/2} & \\
   \boldsymbol{\Theta}_{ji}(-A)^\dag & \mathbf{T}^{3/2}_{ji} & & -\frac{1}{2}\lambda_{\rm SO} {\bf I}^{3/2}
\end{array}
\right)
\label{eq:Hso}
\end{equation}
where ${\bf I}^{1/2, 3/2}$ are the identity matrices for the $j_{\rm eff}$ = 1/2 and 3/2 subspaces,
respectively.
The hopping terms within the $j_{\rm eff}$ = 1/2 and 3/2 subspaces, ${\bf T}^{1/2, 3/2}_{ij}$,
are written in terms of the molecular $t_2$ hopping terms in Eq.~(\ref{eq:H12}) such that
\begin{eqnarray}
  {\bf T}^{1/2}_{ij} &=& t^0 {\bf I} + i\boldsymbol{t}^{\rm D}_{ij} \cdot {\bf S}^{1/2} \nonumber \\
  {\bf T}^{3/2}_{ij} &=& t^0 {\bf I} + i\boldsymbol{t}^{\rm D}_{ij} \cdot {\bf S}^{3/2}
  + \boldsymbol{t}^{\rm Q}_{ij} \cdot \mathbf{\Gamma}  \nonumber
\end{eqnarray}
\begin{eqnarray}
~~~~~~~~~~ {\rm where} ~~~~~~  t^0 &=&  \frac{1}{3} \left( S_1 + S_2 + S_3 \right), \nonumber \\
  \boldsymbol{t}^{\rm D} &=& -\frac{2}{3} \left(A_1, A_2, A_3 \right), \nonumber \\
  \boldsymbol{t}^{\rm Q} &=& -\frac{1}{\sqrt{3}} \left(S_4, S_5, S_6, \frac{S_2-S_3}{2},
                            \frac{2S_1-S_2-S_3}{2\sqrt{3}}  \right), \nonumber
                            \label{eq:jeffhop}
\end{eqnarray}
and ${\bf S}^{1/2}$ and ${\bf S}^{3/2}$ are the pseudospin operators of the $j_{\rm eff}$ = 1/2 and 3/2 states,
respectively. $\boldsymbol{t}^{\rm D}$ couples to ${\bf S}^{1/2,3/2}$ and can be interpreted as the effective
dipolar fields on the hopping electrons, of which directions are shown 
in Fig.~4 {\bf c} in the main text.
For the $j_{\rm eff}$ = 3/2 states, additional quadrupolar fields manifested as the Dirac Gamma matrices
$\boldsymbol{\Gamma}\equiv(\Gamma_1,\Gamma_2,\Gamma_3,\Gamma_4,\Gamma_5)$ couple to the hopping electron,
where the Dirac matrices are defined as$^{5}$ ~$\Gamma_1 = \sigma^z \otimes \sigma^y,
~\Gamma_2 = \sigma^z \otimes \sigma^x,~\Gamma_3 = \sigma^y \otimes {\bf I}^{1/2},
~\Gamma_4 = \sigma^x \otimes {\bf I}^{1/2},~\Gamma_5 = \sigma^z \otimes \sigma^z.$
Note that the Dirac Gamma matrices can be represented in terms of ${\bf S}^{3/2}$ such that
\begin{equation}
\boldsymbol{\Gamma} = \left(
\sqrt{3}\{ S^{3/2}_y, S^{3/2}_z\},
~\sqrt{3}\{ S^{3/2}_z, S^{3/2}_x\},
~\sqrt{3}\{ S^{3/2}_x, S^{3/2}_y\},
~\frac{1}{\sqrt{3}} \left[ (S^{3/2}_x)^2 - (S^{3/2}_y)^2 \right],
~(S^{3/2}_z)^2
\right). \nonumber
\end{equation}
The inter-orbital hopping term $\boldsymbol{\Theta}_{ij}$ is given as
\begin{equation}
\boldsymbol{\Theta}_{ij} \equiv \left(
\begin{array}{cccc}
\frac{ (S_4 + A_1) - i(S_5 + A_2)   }{\sqrt{6}} &
\frac{ (2S_1 - S_2 - S_3) + 2iA_3   }{3\sqrt{2}} &
\frac{ (3S_4 - A_1) + i(3S_5 - A_2) }{3\sqrt{2}} &
\frac{ (S_2 - S_3) + 2iS_6          }{\sqrt{6}} \\
\frac{ (-S_2 + S_3) + 2iS_6         }{\sqrt{6}} &
\frac{ -(3S_4-A_1) + i(3S_5 - A_2)  }{3\sqrt{2}} &
\frac{ (2S_1-S_2-S_3) - 2iA_3       }{3\sqrt{2}} &
\frac{ (S_4 + A_1) + i(S_5 + A_2)   }{\sqrt{6}} \nonumber
\end{array}
\right).
\end{equation}
Note that, from the hermiticity, $\boldsymbol{\Theta}_{ij}(-A) = \boldsymbol{\Theta}_{ji}$.\\

\noindent
\textbf{Supplementary Note 3: Block-diagonalization of $\boldsymbol{j}_{\rm eff}$-based Effective Hamiltonian.~}
Rearranging the Hamiltonian in Eq.~(\ref{eq:Hso}) in terms of the $j_{\rm eff}$ = 1/2 and 3/2
subspaces yields
\begin{equation}
\mathcal{H}^{j_{\rm eff}} = \left(
\begin{array}{cc|cc}
   +\lambda_{\rm SO}  {\bf I}^{1/2} & \mathbf{T}^{1/2}_{ij} & & \boldsymbol{\Theta}_{ij}   \\
 \mathbf{T}^{1/2}_{ji}  & +\lambda_{\rm SO} {\bf I}^{1/2}   & \boldsymbol{\Theta}_{ji} &   \\ \hline
  & \boldsymbol{\Theta}_{ij}(-A)^\dag &   -\frac{1}{2}\lambda_{\rm SO} {\bf I}^{3/2} & \mathbf{T}^{3/2}_{ij} \\
   \boldsymbol{\Theta}_{ji}(-A)^\dag & & \mathbf{T}^{3/2}_{ji} & -\frac{1}{2}\lambda_{\rm SO} {\bf I}^{3/2}
\end{array}
\right).
\label{eq:Hso2}
\end{equation}
As the energy splitting between the $j_{\rm eff}$ = 1/2 and 3/2 states, $\frac{3}{2}\lambda_{\rm SO}$,
is large compared to the inter-orbital hopping terms $\boldsymbol{\Theta}$,
we can block-diagonalize $\mathcal{H}^{j_{\rm eff}}$ into the $j_{\rm eff}$ = 1/2 and 3/2 subspaces.
The correction to the $j_{\rm eff}$ = 1/2 on-site terms are as follows:
{\small
\begin{align}
\Delta \mathcal{H}^{j_{\rm eff}}_{11} &=
   \frac{2}{3\lambda_{\rm SO}} \sum^{12}_{n=5} \mathcal{H}^{j_{\rm eff}}_{1n} \mathcal{H}^{j_{\rm eff}}_{n1} \nonumber \\
   &= \frac{4}{27\lambda_{\rm SO}} \left[
     (S^2_1 + S^2_2 + S^2_3 - S_1S_2 - S_2S_3 - S_3S_1) + 3(S^2_4 + S^2_5 + S^2_6) + (A^2_1 + A^2_2 + A^2_3)
     \right] \nonumber \\
\Delta \mathcal{H}^{j_{\rm eff}}_{12}
    &= \frac{2}{3\lambda_{\rm SO}} \sum^{12}_{n=5} \mathcal{H}^{j_{\rm eff}}_{1n} \mathcal{H}^{j_{\rm eff}}_{n2} \nonumber \\
    &= \frac{2}{3\lambda_{\rm SO}} \Big\{
       \frac{1}{6} \left[ (S_5 - A_2) + i(S_4 + A_1) \right]
                   \left[ (S_2 - A_3) - 2iS_6 \right] \nonumber \\
    & ~~~~~~~~~~~~ +
       \frac{1}{18}\left[ (2S_1-S_2-S_3) + 2iA_3 \right]
                   \left[ (3S_5+A_2) - i(3S_4-A_1) \right] \nonumber \\
    & ~~~~~~~~~~~~ +
       \frac{1}{18}\left[ (2S_1-S_2-S_3) + 2iA_3 \right]
                   \left[-(3S_5+A_2) + i(3S_4-A_1) \right] \nonumber \\
    & ~~~~~~~~~~~~ +
       \frac{1}{6} \left[ (S_5 - A_2) + i(S_4 + A_1) \right]
                   \left[(-S_2 + A_3) + 2iS_6 \right] \Big\} = 0  \nonumber
\end{align}
}
Straightforward calculations for other diagonal terms yield
$\Delta \mathcal{H}^{j_{\rm eff}}_{11} = \Delta \mathcal{H}^{j_{\rm eff}}_{22} =
 \Delta \mathcal{H}^{j_{\rm eff}}_{33} = \Delta \mathcal{H}^{j_{\rm eff}}_{44}$,
and the off-diagonal terms vanish
($\Delta \mathcal{H}^{j_{\rm eff}}_{12} =
\Delta \mathcal{H}^{j_{\rm eff}}_{21} =
\Delta \mathcal{H}^{j_{\rm eff}}_{34} =
\Delta \mathcal{H}^{j_{\rm eff}}_{43} = 0$).
For the 4$d$ transition metal compounds, especially for GaMo$_4$S$_8$,
the on-site energy shift from this inter-subspace mixing is less than 20~meV,
which is an order-of-magnitude smaller than the the SOC splitting.
In the 5$d$ transition metal compounds, the correction becomes negligible.

Similarly, the corrections to the hopping term, ${\bf T}^{1/2}_{ij}$, are as follows:
{\small
\begin{align}
\Delta \mathcal{H}^{j_{\rm eff}}_{13}
    &= \frac{2}{3\lambda_{\rm SO}} \sum^{12}_{n=5} \mathcal{H}^{j_{\rm eff}}_{1n} \mathcal{H}^{j_{\rm eff}}_{n3} \nonumber \\
    &= \frac{2}{3\lambda_{\rm SO}} \left[
        \sum^{8}_{n=5}  \underbrace{\mathcal{H}^{j_{\rm eff}}_{1n}}_{=0} \mathcal{H}^{j_{\rm eff}}_{n3}  +
        \sum^{12}_{n=9} \mathcal{H}^{j_{\rm eff}}_{1n} \underbrace{\mathcal{H}^{j_{\rm eff}}_{n3}}_{=0}  \right] = 0 \nonumber \\
\boldsymbol{\Rightarrow} \Delta \mathcal{H}^{j_{\rm eff}}_{13} &=
\Delta \mathcal{H}^{j_{\rm eff}}_{14} =
\Delta \mathcal{H}^{j_{\rm eff}}_{23} =
\Delta \mathcal{H}^{j_{\rm eff}}_{24} =
\Delta \mathcal{H}^{j_{\rm eff}}_{31} =
\Delta \mathcal{H}^{j_{\rm eff}}_{41} =
\Delta \mathcal{H}^{j_{\rm eff}}_{32} =
\Delta \mathcal{H}^{j_{\rm eff}}_{42} = 0. \nonumber
\end{align}
}
Since the second-order corrections to the hopping elements in the $j_{\rm eff}$ = 1/2 block
vanish, the mixing between $j_{\rm eff}$ = 1/2 and
3/2 blocks through the hopping terms are suppressed in these lacunar spinel compounds.
Consequently, the effective Hamiltonian for the lacunar spinel compounds can be written as
\begin{equation}
\mathcal{H}_{\rm eff} \simeq \mathcal{H}^{1/2} \oplus \mathcal{H}^{3/2}.
\end{equation}\\

\noindent
\textbf{Supplementary Note 4: DFT+SOC+$U$ Results}

\textit{4.A. Calculation Details.~}
In this subsection, we explain the choice of initial magnetic configurations and the range of
$U_{\rm eff}$ values we used in DFT+SOC+$U$ calculations. Three different initial magnetic
configurations within the $M_4$ cluster --- the all-in-all-out, the 2-in-2-out, and the collinear
order, as shown in Supplementary Figure 3 (a), (b), and (c), respectively
--- are tried to detect the non-collinear order.
For the magnetic order between the molecular moments on the neighboring $M_4$ clusters,
ferromagnetic (FM) and antiferromagnetic (AFM) orders are considered.
A doubled unit cell with two formula units is used to incorporate the AFM order.
In order to choose reasonable $U_{\rm eff}$ values, we referred to the work of \c{S}a\c{s}\i o\u{g}lu
and co-workers, where the $U$ and $J$ values for transition metals are evaluated as functions of $d$
orbital occupation from the constrained RPA calculations$^{6}$;
$U_{\rm eff}$ values are estimated around 1.0 $\sim$ 2.0~eV for $d^3$ and $d^4$ configurations of
4$d$ and 5$d$ transition metal atoms. Taking account of the small $d$ orbital occupations --- $d^{1.75}$
for $M$ = Nb/Ta and $d^{2.75}$ for Mo/W --- and spatially extended molecular $t_2$ orbitals,
we suppose that the reasonable value of $U_{\rm eff}$ does not exceed 2.5~eV and 2.0~eV for
the 4$d$ and 5$d$ compounds, respectively. Here, we use $1.5 \leq U_{\rm eff} \leq 2.5$~eV and
$1.0 \leq U_{\rm eff} \leq 2.0$~eV for the 4$d$ and 5$d$ lacunar spinel compounds, respectively.

\textit{4.B. $j_{\rm eff}$-projected Bands and Density of States of FM Configuration.~}
Supplementary Figure 4 shows the $j_{\rm eff}$-projected band structures and PDOS of
GaTa$_4$Se$_4$Te$_4$, GaW$_4$Se$_4$Te$_4$, GaNb$_4$Se$_8$, and GaMo$_4$Se$_8$ in FM order.
The same $U_{\rm eff}$ values with the AFM calculations are used: 2.0 and 2.5~eV for 5$d$ and 4$d$ compounds,
respectively. In common with the AFM results shown in Fig.~3 in the main text, the gap opening and
the robust $j_{\rm eff}$ character are seen in the low-energy spectrum of all the compounds.
The magnetic moments on the four corners of the $M_4$ cluster are collinear for $M$ = Ta, W, Mo compounds,
while in GaNb$_4$Se$_8$ a non-collinear order develops.

\textit{4.C. Molecular $t_2$ Character and Collinearity within $M_4$ Cluster.~}
Once the low-energy electronic degrees of freedom are perfectly characterized by the molecular $t_2$ states,
the angular momenta at the four corners of the $M_4$ cluster should be collinear.
This is purely owing to the nature of the molecular $t_2$ states,
where their orbital components are identical at each transition metal site.
More specifically, any wavefunction written as a linear combination of the molecular $t_2$ states reads
\begin{align}
\vert \psi \rangle
&= \sum^3_{\alpha = 1} \sum_{\sigma = \uparrow\downarrow}
       c^{~\sigma}_\alpha \vert s_z = \sigma \rangle \vert D_\alpha \rangle \nonumber \\
&= \frac{1}{2} \sum^4_{i=1} \left(
   \sum^3_{\alpha = 1} \sum_{\sigma = \uparrow\downarrow}
      c^{~\sigma}_\alpha \vert s_z = \sigma \rangle
   \vert d^{~i}_\alpha \rangle
   \right),
\end{align}
where $i=1,\cdots, 4$ are the $M_4$ corner index, and $\alpha$ denotes the orbital index $xy$, $yz$, $xz$.
Regardless of the site index $i$, spin and orbital components in Eq.~(7) are the same;
in other words, the coefficient $c^{~\sigma}_\alpha$ does not have the site index $i$.
The implication of this result is simple; as long as the molecular $t_2$ states are perfectly isolated
near the Fermi level, magnetic moments at the four corners of the $M_4$ cluster are collinear and
behave as a single moment. The only way to introduce non-collinear order is to make a mixture of
molecular $t_2$ states with other molecular states.

\textit{4.D. Magnetic Order in DFT+SOC+$U$ Calculations.~}
In this subsection, we discuss the magnetic order from the DFT+SOC+$U$ calculations and their $U_{\rm eff}$
dependence. Detailed results --- spin/orbital moments within the $M_4$ cluster, and the relative energy
between the FM and AFM configurations --- are tabulated in Supplementary Table 2-5.

For the $j_{\rm eff}$ = 1/2 compounds ($M$ = Mo, W), collinear orders within the $M_4$ cluster are
observed both in the FM and AFM states in the whole range of $U_{\rm eff}$ values we considered.
Supplementary Figure 5 (a) and (b) show the magnetic moments of GaW$_4$Se$_4$Te$_4$ and
GaMo$_4$Se$_8$ in the FM configuration at $U_{\rm eff}$ = 2.0 and 2.5~eV, respectively.
Spin and orbital angular momentum (SAM and OAM) align parallel to each other owing to the $j_{\rm eff}$ = 1/2
character. The total sum of each moment within the $M_4$ cluster is quite close to the ideal 
$j_{\rm eff}$ = 1/2 moment (1/6 and 2/3 $\mu_B$ for SAM and OAM, respectively) with small
$U_{\rm eff}$ dependence. These reflect the nature of the pure $j_{\rm eff}$ = 1/2 character of the
unoccupied upper Hubbard bands in these compounds. The AFM and the FM states are nearly degenerate;
the energy difference is smaller than 0.3~meV for $U_{\rm eff} \geq 1.5$~eV.
This might imply the competing anisotropic exchange interactions with Heisenberg terms
as mentioned in the main text, which needs more elaborate investigations on magnetism.

For GaTa$_4$Se$_4$Te$_4$, the system changes from a paramagnetic metal to a magnetic insulator
in between $U_{\rm eff}$ = 1.0 and 1.5~eV. Supplementary Figure 5 (c) shows the magnetic moments of
GaTa$_4$Se$_4$Te$_4$ in the FM configuration at $U_{\rm eff}$ = 2.0~eV, where the large OAM with
collinear order dominates over the small canted SAM. SAM and OAM at each of the $M_4$ corners tend
to align antiparallel to each other owing to the $j_{\rm eff}$ = 3/2 character. Contrary to the
$j_{\rm eff}$ = 1/2 systems, the FM configuration becomes more stable than AFM by 6.5~meV per
formula unit at $U_{\rm eff}$ = 2.0~eV.

In GaNb$_4$Se$_8$, the system turns from a paramagnetic metal to a non-collinear ordered insulator for $U_{\rm eff}
\geq$ 2.0~eV; Supplementary Figure 5 (d) shows the magnetic order at $U_{\rm eff}$ = 2.5~eV in the FM calculation.
Such non-collinear orders can be attributed to the mixing between the molecular $t_2$ and the molecular $e$ states,
which is shown in Supplementary Figure 6 depicting the PDOS of GaNb$_4$Se$_8$ at $U_{\rm eff}$ = 2.5~eV.
The AFM calculations, which are energetically more stable than FM results,
show different non-collinear ordering compared to the FM results.
From these results, one can say that GaNb$_4$Se$_8$ shows weaker molecular $t_2$ and $j_{\rm eff}$ character
compared to the other compounds in the presence of large electron correlations.
Still, the molecular $j_{\rm eff}$ = 3/2 character prevails in the low-energy spectrum of GaNb$_4$Se$_8$ as shown
in Supplementary Figure 6.
The competition between the electron correlations and the molecular nature as well as the atomic SOC
may induce complicated internal structures of SAM and OAM within the $M_4$ cluster,
which calls for further studies. \\

\noindent
\textbf{Supplementary Note 5: $\boldsymbol{j}_{\rm eff}$ = 1/2 Spin Hamiltonian.~}
The detailed expression for the $j_{\rm eff}$ = 1/2 spin Hamiltonian in the strong coupling regime
is as follows$^{7}$:
\begin{eqnarray}
  \mathcal{H}^{1/2}_{\rm spin} &=& \sum_{\langle ij \rangle} \left[
  {\rm J}{\bf s}_i\cdot{\bf s}_j +
  {\bf D}_{ij} \cdot \left( {\bf s}_i\times{\bf s}_j \right) +
  {\bf s}_i\cdot{\bf A}_{ij}\cdot{\bf s}_j
  \right]
  \\
{\rm with} ~~~~~~~  {\rm J}     &=& \frac{4}{U} t^2_0 \nonumber \\
  {\bf D}_{ij} &=& \frac{4}{U} t_0 \left( \boldsymbol{t}^{\rm D}_{ij} - \boldsymbol{t}^{\rm D}_{ji} \right) \nonumber \\
  {\bf A}_{ij} &=& -\frac{4}{U} \left(
  \boldsymbol{t}^{\rm D}_{ji}\otimes\boldsymbol{t}^{\rm D}_{ij} + \boldsymbol{t}^{\rm D}_{ij}\otimes\boldsymbol{t}^{\rm D}_{ji}
  - \boldsymbol{t}^{\rm D}_{ij}\cdot\boldsymbol{t}^{\rm D}_{ji}
  \right) \nonumber
\end{eqnarray}
where $\otimes$ denotes the outer product of two vectors.
Note that the Dzyaloshinskii-Moriya vector ${\bf D}_{ij}$ is proportional to
$\boldsymbol{t}^{\rm D}_{ij}$ since $\boldsymbol{t}^{\rm D}_{ij} = - \boldsymbol{t}^{\rm D}_{ji}$.\\

\clearpage

\noindent
\textbf{\Large{Supplementary References}}\\

{\small
\begin{correct}
$^{1}$ C. H. Kim, H.-S. Kim, H. Jeong, H. Jin, and J. Yu,
Topological quantum phase transition in 5$d$ transition metal oxide Na$_2$IrO$_3$,
\emph{Physical Review Letters} {\bf 108}, 106401 (2012).\\
$^{2}$ H.-S. Kim, C. H. Kim, H. Jeong, H. Jin, and J. Yu, 
Strain-induced topological insulator phase and effective magnetic interactions in Li$_2$IrO$_3$,
\emph{Physical Review B} {\bf 87}, 165117 (2013).\\
$^{3}$ A. Camjayi, R. Weht, and M. Rozenberg, 
Localised Wannier orbital basis for the Mott insulators GaV$_4$S$_8$ and GaTa$_4$Se$_8$,
\emph{Europhysics Letters} {\bf 100}, 57004 (2012).\\
$^{4}$ B. J. Kim, H. Jin, S. J. Moon, J. Y. Kim, B. G. Park, C. S. Leem, J. Yu, T. W. Noh, C. Kim, S. J. Oh, J. H. Park, V. Durairaj, G. Cao, and E. Rotenberg, 
Novel $J_{\rm eff}$=1/2 Mott state induced by relativistic spin-orbit coupling in Sr$_2$IrO$_4$,
\emph{Physical Review Letters} {\bf 101}, 076402 (2008).\\
$^{5}$ S. Murakami, N. Nagaosa, and S.-C. Zhang, 
SU(2) non-Abelian holonomy and dissipationless spin current in semiconductors,
\emph{Physical Review B} {\bf 69}, 235206 (2004).\\
$^{6}$ E. \c{S}a\c{s}\i o\u{g}lu, C. Friedrich, and S. Bl{\"u}gel, 
Effective Coulomb interaction in transition metals from constrained random-phase approximation,
\emph{Physical Review B} {\bf 83}, 121101(R) (2011).\\
$^{7}$ T. Micklitz and M. R. Norman, 
Spin Hamiltonian of hyper-kagome Na$_{4}$Ir$_{3}$O$_{8}$,
\emph{Physical Review B} {\bf 81}, 174417 (2010).
\end{correct}
}

\end{document}